\documentclass[%
 aip,
 amsmath,amssymb,
 reprint,%
]{revtex4-1}

\usepackage{graphicx}
\usepackage{dcolumn}
\usepackage{bm}

\usepackage[utf8]{inputenc}
\usepackage[T1]{fontenc}
\usepackage{mathptmx}
\usepackage{etoolbox}
\usepackage{verbatim}

\newcommand*{\rttensor}[1]{\underline{\underline{#1}}}

\makeatletter
\def\@email#1#2{%
 \endgroup
 \patchcmd{\titleblock@produce}
  {\frontmatter@RRAPformat}
  {\frontmatter@RRAPformat{\produce@RRAP{*#1\href{mailto:#2}{#2}}}\frontmatter@RRAPformat}
  {}{}
}%
\makeatother
\begin{document}

\preprint{AIP/123-QED}

\title[Chapman-Enskog derivation of multicomponent Navier-Stokes equations]{Chapman-Enskog derivation of multicomponent Navier-Stokes equations}
\author{Philippe Arnault}
 \email{philippe.arnault@cea.fr.}
\author{S\'ebastien Guisset}%
 \email{sebastien.guisset@cea.fr.}
\affiliation{ 
CEA, DAM, DIF, 91297 Arpajon, France
}%


\date{\today}

\begin{abstract}
There are several reasons to extend the presentation of Navier-Stokes equations to multicomponent systems. Many technological applications are based on physical phenomena that are present neither in pure elements nor in binary mixtures. Whereas Fourier's law must already be generalized in binaries, it is only with more than two components that Fick's law breaks down in its simple form. The emergence of dissipative phenomena affects also the inertial confinement fusion configurations, designed as prototypes for the future fusion nuclear plants hopefully replacing the fission ones. 

This important topic can be described in much simpler terms than in many textbooks since the publication of the formalism put forward recently by Snider in \textit{Phys. Rev. E} \textbf{82}, 051201 (2010). In a very natural way, it replaces the linearly dependent atomic fractions by the independent set of partial densities. Then, the Chapman-Enskog procedure is hardly more complicated for multicomponent mixtures than for pure elements. Moreover, the recent proposal of a convergent kinetic equation by Baalrud and Daligault in \textit{Phys. Plasmas} \textbf{26}, 082106 (2019) demonstrates that Boltzmann equation with the potential of mean force is a far better choice in situations close to equilibrium, as described by the Navier-Stokes equations, than Landau or Lenard-Balescu equations.

In our comprehensive presentation, we emphasize the physical arguments behind Chapman-Enskog derivation and keep the mathematics as simple as possible. This excludes as a technical non-essential aspect the solution of the linearized Boltzmann equation through an expansion in Hermite polynomials. We discuss the link with the second principle of Thermodynamics of entropy increase, and what can be learned from this exposition.
\end{abstract}

\maketitle

%

\section{\label{sec:level1}Introduction}

The Navier-Stokes equations \cite{Landau1992} describe the hydrodynamic evolution in time and space of fluids, with ubiquitous applications in Nature and Technology. Of particular interest are the experiments performed on large laser facilities, to probe extreme states of matter such as encounted in Astrophysics \cite{Drake2018}, or to setup configurations for inertial confinement fusion (ICF) \cite{Lindl2004}. ICF is an emerging technology aimed at providing the next generation of nuclear plants. It provides fusion energy from the implosion of a capsule filled with deuterium and tritum. In this context, there is a renewed interest in kinetic theory since various conditions are met from hydrodynamic-like to strongly out-of-equilibrium phenomena.  

It is a new circumstance that multicomponent diffusion in weakly coupled plasmas must be considered as a part of the hydrodynamic simulation of the ICF capsule implosions \cite{Vold2019,Mackay2020,Stanton2021}. Multicomponent diffusion is also an essential ingredient of many technological applications, reviewed by Krishna \cite{Krishna2019}, where counter-intuitive phenomena occur  as the osmotic diffusion where the presence of a third component leads to an uphill diffusion between the other two components against their concentration gradient. 

The Navier-Stokes equations can be derived from the Thermodynamics of irreversible processes \cite{Groot1984} under two principal assumptions: first that the system be close to thermodynamical equilibrium, second that any gradient of thermodynamic quantity be small. Then, the gradients lead to dissipative phenomena acting against them, that appear as fluxes of mass, momentum, and energy. These fluxes are proportional to the gradients and the coefficients of proportionality are the transport coefficients. Important properties of symmetry of the transport coefficients are obtained from the second principle of Thermodynamics about the increase of entropy of an isolated system. Nevertheless, this route towards the Navier-Stokes equations does not provide any criteria of thermodynamic equilibrium nor does it provide recipes to compute the transport coefficients.

The kinetic theory is another route towards the Navier-Stokes equations that draws a relationship between microscopic events and macroscopic behaviors (see Ref.\,\cite{Ferziger1972,Zhdanov2002} for instance). This leads naturally to a criteria of equilibrium and to a prescription to compute the transport coefficients. Then, the symmetry of the transport coefficients is warranted by construction. Moreover, the second principle of Thermodynamics becomes a consequence of the theory. However, all this is at the cost of a restriction to only deal with systems where the interaction energy between the particles is much weaker than their kinetic energy.

The domain of validity of kinetic theory encompasses the situations of dilute neutral gas at low density $n$, where there is rarely more than two particles within the range $\sigma$ of the interaction potential $V(r)$, so that $n\,\sigma^3 \ll 1$, and the situations of weakly coupled plasmas at high temperature $T$, where the effect of the multiple collisions of a particle with the others can be summed in pairs, since the strength of the potential $V_0$ at typical interparticle distances is far less that the mean kinetic energy, ${V_0}/{kT} \ll 1$ ($k$ is the Boltzmann constant) \cite{Nicholson1983, Baalrud2019}.

The cornerstone of the derivation of the Navier-Stokes equations using kinetic theory is an expansion of the equations  \textit{w.r.t.} a small parameter: the \emph{Knudsen number}, defined later. This permits to develop analytically the foundations of the theory, and to obtain exact results within a controlled validity domain. Unfortunately, the mathematical apparatus of kinetic theory is quite intricate and it often obscures to the novices the physical principles at the heart of the theory. Here, we present a derivation of the multicomponent Navier-Stokes equations, that emphasizes the physical ingredients and keep the analytical developments as simple as possible, thanks to Snider's recent proposal for a new treatment of multicomponent diffusion \cite{Snider2010}. Indeed, the complexity inherent in the treatment of multicomponent systems stems from the appearance of a set of $N$ linearly-dependent concentrations $x_i$, where $N$ is the number of components $i$ in the mixture. 
Fortunately, Snider \cite{Snider2010}  proposed to circumvent this difficulty using the  set of the independent partial densities $n_i$ instead of the concentrations $x_i$. 

 Another recent breakthrough in the theory is the recent work of Baalrud and Daligault \cite{Baalrud2019} on a convergent kinetic equation particularly well adapted to systems close to equilibrium. It is Boltzmann equation with the potential of mean force. To find a closure of the BBGKY hierarchy, they exhibited an expansion parameter independent on the range or the strength of the interaction potential. As a result, their kinetic equation applies equally well to neutral gas and plasmas and admits a particularly large validity domain \cite{Kagan2017}.

We start, in Sec.\,\ref{sec:FluidEq}, by explaining how the fluid equations are obtained as the velocity moments of Boltzmann's equations. It is the occasion to emphasize the central role of the conservation equations of mass, momentum, and energy, in separating slow modes of variations associated with the fluid equations from fast modes  associated with the collision integrals. 

In Sec.\,\ref{sec:Transport}, the main steps of the Chapman-Enskog derivation are described: the fluid scaling that results in the appearance of the inverse Knudsen number $\varepsilon$ in Boltzmann's equation; the separation in order of $\varepsilon$ between Euler's and Navier-Stokes equations; Snider's formulation of the driving forces associated with the different gradients of partial densities, velocity, and temperature, that leads to the corresponding general solution of the linearized Boltzmann equations. The Navier-Stokes equations are then obtained substituting this solution into the fluid equations.

We found that the analysis (made in Sec.\,\ref{sec:TransportCoeff}) of the rate of entropy production highlights the emergence of the dissipation mechanisms associated with the transport coefficients. It is also a useful guide to introduce and define them, providing their properties of symmetry with physical insights, especially as concerns the thermal conductivity.

Sec.\,\ref{sec:OneTwoThree} is dedicated to a concrete illustration of the emergence of the dissipative phenomena as the number of components in a mixture increases from pure elements to binary mixtures, and finally ternary mixtures. We examine here how Fourier's and Fick's laws must be generalized in these situations.

For completeness, we provide two appendices on binary collisions (App.\,\ref{app:Collision}), the derivation of Boltzmann's equation, and its properties (App.\,\ref{app:Boltzmann}). Other appendices are also provided that give the missing steps of some derivations.

\section{From Boltzmann kinetic equation to fluid equations}
\label{sec:FluidEq}

\subsection{Boltzmann equation}

The Boltzmann kinetic equation describes the evolution in time $t$, space $\textbf{r}$, and velocity $\textbf{v}$ of the distribution function $f_i(t,\textbf{r},\textbf{v})$. We choose the normalisation of $f_i$ where $f_i(t,\textbf{r},\textbf{v})\, d^3\textbf{r}\, d^3\textbf{v}$ is the number of particles of species $i$ (and mass $m_i$) at time $t$, in the volume element of space between $\textbf{r}$ and $\textbf{r} + d^3\textbf{r}$, and in the volume element of velocity between $\textbf{v}$ and $\textbf{v} + d^3\textbf{v}$.
In the absence of external force, 
the Boltzmann equation then reads
\begin{subequations}
\begin{equation}
\label{Boltmann_eq}
\partial_t f_i +  \textbf{v}_i\cdot \nabla f_i = \sum_j J[f_i,f_j](\textbf{v}_i),
\end{equation}
with the Boltzmann collision integrals (see App.\,\ref{app:CollisionOperator})
\begin{equation}
J[f_i,f_j](\textbf{v}_i) = \int (f_i'f'_j-f_if_j) \, \mathbf{v}_{ij}\, 2 \pi \, b \, d b \, d^3 \textbf{v}_{j}.
\end{equation}
The shortcut notation $f_i'$ stands for $f_i(t,\textbf{r},\textbf{v}_i')$ where $\textbf{v}_i'$ is the velocity of a particle of species $i$ after its binary collision with a particle of species $j$ of initial velocity $\textbf{v}_j$. Actually, the notation $f_i$ embarks also the information about species $i$: its mass $m_i$ and density $n_i$. In the binary collision, $\textbf{v}_{ij} = \textbf{v}_{j} -\textbf{v}_{i}$ corresponds to the initial relative velocity and the variable $b$ is the impact parameter defined as the minimal distance between the two particles if they do not interact and move in straight lines.
Actually, the way this equation is written is highly symbolic since the velocities after collision are non-trivial functions of the initial velocities (see App.\,\ref{app:Collision})
\begin{equation}
\textbf{v}_i' = \dfrac{m_j}{m_i+m_j} \textbf{v}'_{ij} + \textbf{V}, 
\end{equation} 
with the center of mass velocity
\begin{equation}
    \textbf{V} = \dfrac{m_i \textbf{v}_{i}+m_j \textbf{v}_{j}}{m_i+m_j},
\end{equation}
and the relative velocity after collision
\begin{equation}
    \textbf{v}'_{ij} = v_{ij} \left[ \cos(\chi) \, \textbf{e}_x + \sin(\chi)\,\textbf{e}_y \right],
\end{equation}
for a cartesian system where the $x$-axis is along the initial relative velocity, $\textbf{e}_x = \textbf{v}_{ij} / v_{ij}$, and $\textbf{e}_y = \textbf{e}_z \times \textbf{e}_x$ with $\textbf{e}_z = (\textbf{v}_i \times \textbf{v}_j)/ v_i v_j $, a unit vector perpendicular to the trajectory plane. Finally, the deflection angle $\chi$ is related to the impact parameter $b$, the relative velocity $v_{ij}$, and the pair interaction potential $E_P(r)$ by
\begin{equation}
    \chi = \chi(b,v_{ij},E_P) = \pi - 2  \int_{r_0}^\infty \dfrac{1}{r^2 }\dfrac{b dr}{\sqrt{1 - \dfrac{b^2}{r^2} - W(r)}},
\end{equation}
where $r_0$ is the turning point (the largest root of the denominator in the integrand) and
\begin{equation}
 W(r)= \dfrac{E_P(r)}{\dfrac{1}{2} m_{ij} v_{ij}^2},
\end{equation} 
with $m_{ij} = m_i m_j/(m_i+m_j)$, the reduced mass.
\end{subequations} 

\subsection{Collisional invariants}
\label{sec:Invariants}

It is crucial to realize that the collision operator $J[f_i,f_j](\textbf{v}_i)$ of the Boltzmann kinetic equation represents the net effect between the collisions with species $j$ depleting the distribution function $f_i$ of species $i$ at a given velocity $\textbf{v}_i$ and the inverse collisions replenishing the distribution (see App.\,\ref{app:CollisionOperator}). Since the hydrodynamic equations are obtained as the velocity moments of Boltzmann's equation, we shall see that the only way to get rid of the collision terms is to sum each moment equation over all species in the mixture. By doing so, the conservation of mass, momentum, and energy applies whatever the distribution functions and the resulting fluid equations do not involve collisional (friction) terms. The only link to the distribution functions appears through the expressions of the transport coefficients, which constitute the closures of the fluid equations.

Actually, this comes from the fundamental properties of a collisional operator with respect to the collisional invariants corresponding to the conservation of mass, momentum, and energy. Indeed, summing over all the collisions involving species $i$ directly leads to 
\begin{subequations}
\begin{equation}
\label{eq:invariant_mass}
\sum_j \int J[f_i,f_j](\textbf{v}_i) ~ d^3\textbf{v}_{i} = 0,
\end{equation}
so that no particles are created nor annihilated. This is verified by Boltzmann's collision operator for each separated value of $j$ using the relation Eq.\eqref{eq:invariant1} with $K(\textbf{v}) = m$.
In addition by summing over all the collisions one gets that 
\begin{equation}
\label{eq:invariant_momentum}
\sum_{i,j} \int m_i \textbf{v}_{i} \, J[f_i,f_j](\textbf{v}_i) ~ d^3\textbf{v}_{i} = 0,
\end{equation}
and 
\begin{equation}
\label{eq:invariant_energy}
\sum_{i,j} \int \dfrac{1}{2} m_i {v}^2_{i} \, J[f_i,f_j](\textbf{v}_i) ~ d^3\textbf{v}_{i} = 0,
\end{equation}
\end{subequations}
include the momentum and the energy conservation equations for each collision. This is verified by Boltzmann's collision operator using the relation Eq.\eqref{eq:invariant2} with $K(\textbf{v}) = m \textbf{v}$ and $K(\textbf{v}) = \frac{1}{2}m\textbf{v}^2$ respectively. Consequently, $m$, $m \textbf{v}$ and $\frac{1}{2}m\textbf{v}^2$ are collisional invariants of the Boltzmann integrals. These properties will be used extensively in the next section. We also recall Eq.\eqref{eq:invariant2} for later use
\begin{align}
\label{eq:invariant2bis}
\sum_{i, j} &\int  K(\textbf{v}_{i}) \, J[f_i,f_j](\textbf{v}_{i}) \, d^3 \textbf{v}_{i}  \\
&= -\dfrac{1}{4}  \sum_{i, j} \int (K_i' +K_j'-K_i -K_j) \, (f_i' f_j'-f_i f_j)\nonumber \\
& \hspace{3cm} \times \,v_{ij}\,2 \pi b\,db\, d^3 \textbf{v}_{i} d^3 \textbf{v}_{j}.\nonumber
\end{align}

\subsection{Fluid equations}
\label{Fluid}	
First, it is worth recalling that the space and velocity variables of the Boltzmann equation are independent. Therefore, the advection term reads
\begin{equation}
\partial_t f_i +  \textbf{v}_i\cdot \nabla f_i = \partial_t f_i + \nabla\cdot( \textbf{v}_i f_i).
\end{equation}
The integration of the Boltzmann equation over velocity, exploiting the global conservation property Eq.\,\eqref{eq:invariant_mass} of the collisional invariants, gives rise to the following \emph{species mass conservation} equation
\begin{subequations}
\label{eq:mass_eq}
\begin{equation}
\partial_t \rho_i + \nabla \cdot \left( \rho_i \textbf{u} \right) + \nabla \cdot \left( \rho_i \textbf{U}_i \right)  = 0,
\end{equation} 
describing the space-time evolution of the partial density $\rho_i$, defined by
\begin{equation}
\rho_i = m_i n_i = m_i \int f_i(t, \textbf{r}, \textbf{v})  \, d^3\textbf{v}.
\end{equation} 
The total density $\rho= \sum_i \rho_i = m\,n$ and correspondingly $n = \sum_i n_i$ and $m=\sum_i x_i \, m_i$ with $x_i = n_i/n$.
The fluid velocity reads
\begin{equation}
\textbf{u}= \frac{1}{\rho}\sum_i \rho_i \textbf{u}_i, 
\end{equation}
where the peculiar velocity $\textbf{u}_i$ of species $i$ is defined by 
\begin{equation}
n_i \textbf{u}_i = \int \textbf{v} \, f_i(t, \textbf{r}, \textbf{v}) \, d^3\textbf{v}.
\end{equation} 
We write it $\textbf{U}_i$ in the co-mobile frame
\begin{equation}
\label{eq:MassFlux}
\textbf{U}_i = \textbf{u}_i - \textbf{u} = \dfrac{1}{n_i}\,\int (\textbf{v} -\textbf{u}) \, f_i(t, \textbf{r}, \textbf{v}) \, d^3\textbf{v}.
\end{equation} 
\end{subequations}

The \emph{momentum conservation} equation is obtained by multiplying the Boltzmann equation \eqref{Boltmann_eq} by $m_i \textbf{v}_i$, integrating in velocity, 
and summing over all the species to cancel any contribution from the collisional operators according to Eq.\,\eqref{eq:invariant_momentum} (see Eqs.\eqref{eq:momentum} in App.\,\ref{app:fluidEq})
\begin{subequations}
\label{momentum_eq}
\begin{align}
\nonumber
\sum_i \int  d\textbf{v}_i \, &  m_i \textbf{v}_i \, \times \quad \eqref{Boltmann_eq} \\
& \Leftrightarrow \quad \partial_t (\rho \textbf{u}) + \nabla\cdot\left( \rho \textbf{u} \otimes \textbf{u} + \rttensor{\Pi} \right)=0,
\end{align}
with the pressure tensor 
\begin{equation}
\label{eq:MomentumFlux}
\rttensor{\Pi} = \sum_i m_i \int (\textbf{v}-\textbf{u}) \otimes (\textbf{v}-\textbf{u}) \, f_i(t,\textbf{r},\textbf{v}) \, d^3 \textbf{v}.
 \end{equation}
\end{subequations}
Similarly the \emph{energy conservation} equation is obtained by multiplying the Boltzmann equation \eqref{Boltmann_eq} by $\frac{1}{2} m_i v_i^2$, integrating in velocity, and summing over all the species to cancel any contribution from the collisional operators according to Eq.\,\eqref{eq:invariant_energy} (see Eqs.\,\eqref{eq:energy} in App.\,\ref{app:fluidEq})
\begin{subequations}
\label{energy_eq}
\begin{align}
\nonumber
\sum_i \int & d\textbf{v}_i \, \dfrac{1}{2} m_i v_i^2 \, \times \quad \eqref{Boltmann_eq} \\
& \Leftrightarrow \quad \partial_t E + \nabla \cdot \left( E \,\textbf{u} + \rttensor{\Pi} \cdot \textbf{u} + \textbf{q} \right) = 0.
\end{align}
The energy density $E$ is defined by
\begin{equation}
E = \sum_i \int \dfrac{1}{2} m_i v^2 \, f_i(t,\textbf{r},\textbf{v}) \, d^3\textbf{v} = \dfrac{1}{2} \rho u^2 + \dfrac{3}{2} n\, k T,
\end{equation}
where $T$ is the temperature and $k$ Boltzmann's constant.
The heat flux is defined by 
\begin{equation}
\label{eq:HeatFlux}
\textbf{q} = \sum_i \int \dfrac{1}{2} m_i (\textbf{v}-\textbf{u})^2 (\textbf{v}-\textbf{u}) \, f_i(t,\textbf{r},\textbf{v}) \, d^3 \textbf{v}.
\end{equation}
\end{subequations}
Finally, we point out that the flow variables are  $n_i$, $\textbf{u}$, and $T$, and some closure relations must be provided for the fluxes $\textbf{U}_i,\rttensor{\Pi}, \textbf{q}$ to complete the fluid equations system \eqref{eq:mass_eq}-\eqref{momentum_eq}-\eqref{energy_eq}. The procedure leading to the multicomponent Navier-Stokes closure relations is explained in Sec.\,\ref{sec:Transport}.

\subsection{Euler's closures}
\label{sec:Euler}
The most simple closure relation consists in considering the velocity distribution functions $f_i$ in local thermodynamic equilibrium. In that case, they write under the form of Maxwellian distribution functions $M_i$
\begin{subequations}
\begin{align}
\label{Maxwellian} &M_i(t,\textbf{r},\textbf{v})  =\\
n_i&(t,\textbf{r}) \left( \dfrac{m_i}{2 \pi \,k T(t,\textbf{r})} \right)^{3/2} \exp \left(-\dfrac{m_i\left(\textbf{v}-\textbf{u}(t,\textbf{r})\right)^2}{2\, k T(t,\textbf{r})} \right)
\nonumber
\end{align}
where all the space-time dependence is conveyed by the variables $n_i$, $\textbf{u}$, and $T$.
Inserting relation \eqref{Maxwellian} into the definitions $\textbf{U}_i,\rttensor{\Pi}, \textbf{q}$ one recovers the well-known Euler equations with
\begin{equation}
\textbf{U}_i = \dfrac{1}{n_i}\,\int (\textbf{v} -\textbf{u}) \, M_i(\textbf{v}) \, d^3\textbf{v} = 0,
\end{equation} 
\vspace{-0.5cm}
\begin{equation}
\rttensor{\Pi} = \sum_i m_i \int (\textbf{v}-\textbf{u}) \otimes (\textbf{v}-\textbf{u}) \, M_i(\textbf{v}) \, d^3 \textbf{v} = p \,\rttensor{Id},
 \end{equation}
 \vspace{-0.5cm}
\begin{equation}
\textbf{q} = \sum_i \int \dfrac{1}{2} m_i (\textbf{v}-\textbf{u})^2 (\textbf{v}-\textbf{u}) \, M_i(\textbf{v}) \, d^3 \textbf{v} = 0,
\end{equation}
where $p = n kT$ is the ideal gas pressure.
\end{subequations}

\section{Chapman-Enskog formulation}
\label{sec:Transport}

\subsection{Fluid scaling at small Knudsen}
\label{sec:Knudsen}

The fluid scaling of kinetic equations introduces a small parameter, the Knudsen number $\varepsilon$ representing the ratio between the small space and time scales of the microscopic processes and the large scales of the macroscopic flows. This allows one to develop a perturbation expansion of the velocity distribution function. Eventually, this perturbation development gives rise to the Navier-Stokes (NS) hydrodynamic equations.

From an operational point of view, it is worth getting a procedure to estimate $\varepsilon$ in any given situation in order to monitor its smallness and  to assess the validity of the NS equations. As a first step, we consider that the macroscopic time scale $T_0$ and length scale $L_0$ result from an analysis of the solution of Euler equations. This corresponds to the assumption that $f_i = M_i$. Assume $f$ still varies on macroscopic scales close to equilibrium and introduce dimensionless quantities 
\begin{subequations}
\begin{equation}
\tilde{t} = t / T_0, \hspace{1cm} \tilde{r} = r / L_0, \hspace{1cm} \tilde{v} = v / c_0,
\end{equation}
where $c_0$ is of the order of the sound speed.
Let $\mathcal{N}$ be the number of particles within the volume $L_0^3$
\begin{equation}
\label{eq:N}
n_i = \dfrac{\mathcal{N}}{L_0^3},
\end{equation}
the dimensionless particle distribution function is then defined by
\begin{equation}
\tilde{f_i}(\tilde{t},\tilde{\textbf{r}},\tilde{\textbf{v}}) = \dfrac{L_0^3 \, v_{th}^3}{\mathcal{N}} f_i(t,\textbf{r},\textbf{v}),
\end{equation}
\end{subequations}
where $v_{th}$ is the thermal velocity. It is worth remarking that the sound speed $c_0$ is of the same order as the thermal velocity $v_{th} \propto \sqrt{k T / m_i}$. We shall also assume that it is of the order of $L_0 / T_0$. At contrast, the length scale $\lambda$ and the time scale $\tau$ associated with the collisions between particles are of microscopic nature, far less than their macroscopic counterpart, $L_0$ and $T_0$. Table \ref{bosons} gathers these characteristic scales.

 \begin{table}[h!]
\centering
\caption{Macroscopic and microscopic characteristic scales}
\begin{ruledtabular}
\begin{tabular}{c c c}
		& Macroscopic  & Microscopic  \\
\hline	
Length & Smallest gradient length  & Mean free path \\
$[L]$ & $L_0$ & $\lambda = v_{th} \, \tau$ \\
\\
Time & Smallest time scale & Collision time \\
$[T]$ & $T_0$ & $\tau$ \\
\\
Velocity & Sound speed & Thermal velocity \\
$[L/T]$ & $c_0$ $\approx v_{th}$ & $v_{th}$ \\
\end{tabular}
\end{ruledtabular}
\label{bosons}
\end{table}

\noindent  The microscopic length scale to compare with $L_0$ is the mean free path $\lambda$ between two collisions. However, the only space variable of the Boltzmann collision integrals is the impact parameter $b$, of the order of the maximum impact parameter $b_0$ for the collisions with the smallest deflection of the particles.

For weakly coupled plasmas, the maximum impact parameter $b_0$ is of the order of the Debye length $\lambda_D$  \cite{Nicholson1983,Murillo2004}. This length scale characterizes the screening of a test charge by the unlike charges pilled up around it and the like charges repelled from it. The Coulomb potential $V_C(r) = Q/r$ of the test charge $Q$ is dressed by this shielding cloud to form a Debye-Hückel (DH) potentiel $V_{DH}(r) = Q\, \exp(r/\lambda_D)/r$. This DH potential is the solution of a Poisson-Boltzmann system of equations, linearized \textit{w.r.t.} $V_C(r)/kT$ . This linearization is a good approximation in the validity domain of kinetic theory, often characterized by large values of the parameter $\Lambda = n \lambda_D^3$, which represents the number of charges in a Debye cube. In these conditions, each charge interacts simultaneously with many other charges in a Debye sphere. Most of these collisions gives rise to razing diffusion and the mean free path $\lambda$ is defined as the typical distance where the cumulative effect of the different collisions is associated with a substantial deflection \cite{Nicholson1983}. 

In another equivalent definition, the mean free path $\lambda$ is defined from a collision frequency $\nu_c$ itself defined from a multi-fluid approach to the hydrodynamic equations  \cite{Decoster1998}. When Maxwellian distribution functions $f_i(\mathbf v)$ of species with equal temperature but different mean velocities $\mathbf u_i$ are introduced in the collision integrals, the velocity moments of the kinetic equations give rise to friction terms that can be put under the form $m_i~n_i~\nu_{i j}~(\mathbf u_j - \mathbf u_i)$ to define a collision frequency $\nu_c = \nu_{ij}$ for the interaction between species $i$ and $j$.

In both approaches, the mean free path $\lambda$ is proportional to the Debye length $\lambda_D$ and inversely proportional to the parameter $\Lambda$
\begin{equation}
\label{eq:lambda_P}
\lambda \propto \dfrac{\lambda_D}{\Lambda} = \dfrac{1}{n \lambda_D^2}.
\end{equation}

For dilute gas, the mean free path $\lambda$ depends on both $b_0$ and the density $n_i$. Indeed, consider a tube of length $\lambda$ and radius $b_0$ in front of a particle. In a dilute gas, the hypothesis of binary collision requires that there can be only one particle in this tube, which will undergo a collision. Therefore, 
\begin{equation}
\label{eq:lambda}
n_i \, b_0^2 \, \lambda \sim 1.
\end{equation}
We are now ready to propose the fluid scaling of the Boltzmann equation. The advection term scales as
\begin{subequations}
\begin{equation}
\partial_t f_i + \textbf{v} \cdot \nabla f_i =  \dfrac{\mathcal{N}}{L_0^3 \, v_{th}^3} \left( \dfrac{1}{T_0} \partial_{\tilde t} \tilde f_i + \dfrac{c_0}{L_0} \mathbf{\tilde v} \cdot \tilde \nabla \tilde f_i\right),
\end{equation}
whereas the collision term scales as
\begin{equation}
J[f_i,f_j](\textbf{v}_i) = \left(\dfrac{\mathcal{N}}{L_0^3 \, v_{th}^3}\right)^2 \, v_{th}^4 \, b_0^2 \, ~ \tilde J[\tilde f_i, \tilde f_j].
\end{equation}
 Dropping the tilde notations and equaling $c_0$, $v_{th}$, and $L_0/T_0$, the dimensionless multi-species Boltzmann equation writes
\begin{equation}
\partial_t f_i + \textbf{v} \cdot \nabla f_i = \dfrac{\mathcal{N} b_0^2}{L_0^2}\sum_j J(f_i,f_j),
\end{equation}
The collision operator pre-factor is proportional to the inverse Knudsen number $\varepsilon$. It reads
\begin{equation}
\dfrac{\mathcal{N} b_0^2}{L_0^2} = L_0\, \dfrac{\mathcal{N}}{L_0^3}\, b_0^2 = L_0\, n_i b_0^2= \dfrac{L_0}{\lambda} = \dfrac{1}{\varepsilon},  
\end{equation}
\end{subequations}
using Eqs.\,\eqref{eq:N} and \eqref{eq:lambda_P} or \eqref{eq:lambda}.
$\varepsilon$ is the ratio between the microscopic and macroscopic length scales. It is also the ratio between the microscopic and macroscopic time scales, since $L_0 \propto c_0 T_0$ and $\lambda \propto v_{th} \tau$
\begin{equation}
\varepsilon = \dfrac{\lambda}{L_0} = \dfrac{\tau}{T_0}.
\end{equation}

\subsection{Chapman-Enskog ansatz}

The formulation proposed by Chapman and Enskog \cite{Chapman1970} starts from the Boltzmann equation in the reduced units of the fluid scaling. It then introduces a close relationship between the expansion of the velocity distribution functions according to the order in $\varepsilon$, the Knudsen number, and the hydrodynamic equations : Euler's equations control the leading order, and the NS equations are associated with the next-to-leading order. This important point is often obscured by the complicated mathematical apparatus accompanying the calculation of the transport coefficients, \textit{i.e.} the development of the solution in orthogonal Sonine polynomials. In the following the derivation procedure is kept as simple as possible principally because we do not present the practical calculation of the transport coefficients, but only their expressions as functionals of the solution of the kinetic equations. The final expressions are naturally translated in dimensional units by equating $\varepsilon$ to 1.

The Chapman-Enskog ansatz prescribes to look for a perturbation $\mathcal{\phi}_i(t,\textbf{r},\textbf{v})$ of the particle distribution function  
\begin{equation}
\label{eq:CE_ansatz}
f_i(t,\textbf{r},\textbf{v})= M_i(t,\textbf{r},\textbf{v})\left[1+ \varepsilon \,\mathcal{\phi}_i(t,\textbf{r},\textbf{v})\right],
\end{equation}
keeping the relations between $n_i$, $\textbf{u}$, $T$ and the leading order ($\varepsilon = 0$).
Therefore, the unknown $\mathcal{\phi}_i(t,\textbf{r},\textbf{v})$ must fulfill the following constraints 
\begin{subequations}
\label{eq:ChapmanAnsatz}
\begin{equation} 
\int M_i(\textbf{v}) \, \mathcal{\phi}_i(\textbf{v})\, d^3 \textbf{v} =0,
\end{equation}
\vspace{-0.3cm}
\begin{equation}
\sum_i m_i\int \textbf{v} \, M_i(\textbf{v}) \, \mathcal{\phi}_i(\textbf{v})\, d^3 \textbf{v} =0,
\end{equation}
\vspace{-0.3cm}
\begin{equation}
\sum_i \int \dfrac{m_i}{2} (\textbf{v}-\textbf{u}_i)^2 \, M_i(\textbf{v}) \, \mathcal{\phi}_i(\textbf{v})\, d^3 \textbf{v} =0.
\end{equation}
So we have the following relations giving the partial densities $n_i$, the fluid velocity $\textbf{u}$, and the temperature $T$ as integrals over the leading order, \textit{i.e.} Maxwellians, whatever the order of the development of the distribution functions $f_i$.
\begin{equation} 
n_i(t,\textbf{r}) = \int M_i(t,\textbf{r},\textbf{v})\, d^3\textbf{v},
\end{equation}
\vspace{-0.3cm}
\begin{equation}
\rho \,\textbf{u}(t,\textbf{r}) =  \sum_i m_i \int \textbf{v} M_i(t,\textbf{r},\textbf{v})\, d^3 \textbf{v},
\end{equation}
\vspace{-0.3cm}
\begin{equation}
\dfrac{3}{2}n\,k T(t,\textbf{r}) = \sum_i \int \dfrac{m_i}{2} (\textbf{v}-\textbf{u}_i)^2 M_i(t,\textbf{r},\textbf{v})\, d^3\textbf{v}.
\end{equation}
\end{subequations}
Due to the fluid scaling, there is a shift in $\varepsilon^n$ orders
\begin{equation}
\varepsilon~\Biggl[ \partial_t f_i + \textbf{v}_i \cdot \nabla f_i \Biggr]_{(n)} = \left[\sum_j J[f_i,f_j](\textbf{v}_i)\right]_{(n+1)}
\end{equation}
More precisely at order $n = 0$ ($f_i = M_i$) the following relation is obtained 
\begin{equation}
\sum_j J[M_i,M_j](\textbf{v}_i)=0,
\end{equation}
which is verified since  $M'_i M'_j = M_i M_j$ according to the energy conservation in each binary collision, Eq.\,\eqref{eq:binary_momentum_energy}. At next order $n = 1$, Eq.\,\eqref{eq:CE_ansatz} leads to
\begin{equation}
\label{eq:eq_phi}
\partial_t M_i + \textbf{v}_i \cdot \nabla M_i  = \sum_j I[\phi_i,\phi_j](\textbf{v}_i),
\end{equation}
with now \emph{linear} collision integrals (by using the fact that $M'_i M'_j = M_i M_j$)
\begin{equation}
\label{eq:Boltzamnn_linear}
I[\phi_i,\phi_j](\textbf{v}_i) = \int M_i M_j \, (\phi_i'+\phi_j'-\phi_i-\phi_j) \, v_{ij} \, 2 \pi \, b \, d b \, d^3 \textbf{v}_{j}.
\end{equation}
The left hand side of Eq.\,\eqref{eq:eq_phi} comprises the time derivatives and the spatial gradients of the partial densities $n_i$, the fluid velocity $\textbf{u}$, and the temperature $T$. The time derivatives are reduced to spatial gradients in next section. When the gradients vanish, Eq.\,\eqref{eq:eq_phi} is homogeneous. Clearly, the general solution of this homogeneous equation is a linear combination of the collisional invariants. However, we are looking for perturbations around equilibrium, driven by the gradients. Therefore, we discard the solutions involving the collisional invariants.

\subsection{Snider's approach to driving forces}

In accordance with the shift in $\varepsilon$ order, the left hand side of Eq.\,\eqref{eq:eq_phi} must be evaluated at  order 0. This means, in particular, that the time derivative of the partial densities $n_i$, the fluid velocity $\textbf{u}$, and the temperature $T$ are related to their spatial gradients by Euler equations (see Sec.\,\ref{sec:Euler}). The calculations are eased using the following form
\begin{subequations}
\label{eq:driving}
\begin{equation}
\partial_t M_i + \textbf{v}_i \cdot \nabla M_i  = M_i \left[ \partial_t \log(M_i) + \textbf{v}_i \cdot \nabla  \log(M_i)  \right],
\end{equation}
with
\begin{equation}
\log(M_i) = \log(n_i) - \dfrac{3}{2} \log(T) - \dfrac{m_i (\textbf{v}_i-\textbf{u})^2}{2 k T}+ cste.
\end{equation}
Now by setting $\textbf{c}_i=\textbf{v}_i-\textbf{u}$  one gets the following relation (all the computational details are presented in App.\,\ref{app:Snider})
\begin{align}
\dfrac{1}{M_i}\,(\partial_t M_i + \textbf{v} \cdot \nabla M_i)
&= \left( \dfrac{m_i \textbf{c}^2}{2 k T}- \dfrac{3}{2}-\dfrac{m_i}{m}\right) \, \textbf{c} \cdot \nabla \log T \nonumber\\
&+ \dfrac{m_i}{k T} ~ \textbf{c} \otimes \textbf{c} : \rttensor{S} \\
&+\dfrac{1}{n_i}\,\sum_k \left( \delta_{ik} - \dfrac{m_i n_i}{\rho}\right)\, \textbf{c} \cdot \nabla n_k,\nonumber
\end{align}
where the vectorial symmetry has limited the action of the spatial gradients of the fluid velocity to the traceless symmetric rate-of-shear tensor
\begin{equation}
(\rttensor{S})_{\alpha \beta} = \dfrac{1}{2} \,\left(\nabla_\alpha u_\beta + \nabla_\beta u_\alpha\right) - \dfrac{1}{3} \, (\nabla \cdot \textbf{u}) ~ \delta_{\alpha \beta}.
\end{equation}
\end{subequations}
It is worth remarking that this expression is the most natural one, as compared to the standard approach where the form of Fick's law is anticipated with the introduction of the gradients of concentration $x_i = n_i/n$ instead of the gradients of $n_i$. However, this usual approach leads to difficulties in the subsequent derivations since the concentrations $x_i$ are linearly dependent. To our knowledge, Snider \cite{Snider2010} was the first to follow the route of independent particle density $n_i$, that we shall develop in this paper for its much simpler framework.

\subsection{Solution of the linearized Boltzmann equation}
\label{sec:solution}

The first argument to put forward when solving the linearized Boltzmann equation is that space and time variables, $\textbf{r}$ and $t$, do not appear explicitly in Eq.\,\eqref{eq:eq_phi} -- \eqref{eq:driving}. The solution depends on them through the macroscopic variables only: $n_i(t,\textbf{r})$, $T(t,\textbf{r})$, and $\textbf{u}(t,\textbf{r})$.

The second argument concerns the velocity variable $\textbf{v}$ of the distribution functions $f_i$, which only appears in the combination $\textbf{c}=\textbf{v}-\textbf{u}$ in Eq.\eqref{eq:driving}. In Eq.\,\eqref{eq:Boltzamnn_linear}, changing $\textbf{v}$ for $\textbf{c}$ just amounts to warrant the Galilean invariance of the binary collisions. As a consequence, the distribution functions $f_i$ depend on velocity through the variable $\textbf{c}$.

Since the gradients of the partial densities $n_i$, of the fluid velocity $\textbf{u}$, and of the temperature $T$ are all independent, it is tempting to look for solutions in each case when only one gradient exists. We shall work out in details the case of the temperature gradient to highlight the main arguments. In this case, the linearized Boltzmann equation reads
\begin{subequations}
\begin{equation}
\label{eq:phiT}
\sum_j I\left[\phi^T_i,\phi^T_j\right](\textbf{c}_i) = M_i\, \left( \dfrac{m_i c_i^2}{2 k T}- \dfrac{3}{2}-\dfrac{m_i}{m}\right) \, \textbf{c}_i \cdot \nabla \log T,
\end{equation}
where we note this peculiar solution for $\phi_i$ as $\phi^T_i$. Actually, the different components of the temperature gradient are also independent, and focusing on one component $\nabla_\alpha \log T$ along the $\alpha = x, y,$ or $z$ direction, it is clear that the solution $\phi^T_i$ must be proportional to this component of the gradient, for all the species. Therefore, the solution is of the form
\begin{equation}
\phi^T_i(\textbf{c}) =  \textbf{K}^T_i(\textbf{c}) \cdot \nabla \log T,
\end{equation}
and the vectorial function $\textbf{K}^T_i$ is solution to
\begin{equation}
\label{eq:KTi}
\sum_j I\left[\textbf{K}^T_i,\textbf{K}^T_j\right](\textbf{c}_i) = M_i\,\left( \dfrac{m_i c_i^2}{2 k T}- \dfrac{3}{2}-\dfrac{m_i}{m}\right) \, \textbf{c}_i.
\end{equation}
Now, the rotational invariance of the Boltzmann collision operator implies that (see App.\,\ref{app:Rot})
\begin{equation}
\textbf{K}^T_i(\textbf c) = - K^T_i(c) ~ \textbf c,
\end{equation}
where the minus sign has been added to anticipates the counteraction of the heat flux  $\textbf{q}$ to the gradient of temperature $T$.
\end{subequations}

Similarly, when only the rate-of-shear tensor $\rttensor S$ exists, the peculiar solutions $\phi^S_i$ are all proportional to it, and of the form
\begin{subequations}
\begin{equation}
\phi^S_i(\textbf{c}) = \rttensor{K}^S_{~i}(\textbf{c}) : \rttensor{S},
\end{equation}
where the tensorial function $\rttensor{K}^S_{~i}$ is solution to 
\begin{equation}
\label{eq:KSi}
\sum_j I\left[\rttensor{K}^S_{~i},\rttensor{K}^S_{~j}\right](\textbf{c}_i) = M_i\,\dfrac{m_i}{k T} \, \textbf{c}_i \otimes \textbf{c}_i,
\end{equation}
and rotational invariance implies that
\begin{equation}
\rttensor{K}^S_{~i}(\textbf c) = - K^S_i(c) ~\textbf{c}_i \otimes \textbf{c}_i,
\end{equation}
where the minus sign has been added to anticipates the counteraction of the momentum flux $\rttensor{\Pi}$ to the gradients of  fluid velocity $\textbf{u}$.
\end{subequations}

When only exists the gradient of a peculiar species density $n_k$, the solutions $\phi^k_i$ are all proportional to it, of the form
\begin{subequations}
\begin{equation}
\phi^k_i(\textbf{c}) = \textbf{K}^k_i(\textbf{c}) \cdot \nabla n_k,
\end{equation}
with the vectorial function $\textbf{K}^k_i$  solution to
\begin{equation}
\label{eq:Kij}
\sum_j I\left[\textbf{K}^k_i,\textbf{K}^k_j\right](\textbf{c}_i) = M_i\,\dfrac{1}{n_i}\, \left( \delta_{ik} - \dfrac{m_i n_i}{\rho}\right)\, \textbf{c}_i,
\end{equation}
and of the form
\begin{equation}
\textbf{K}^k_i(\textbf c) = - K^k_i(c) ~ \textbf c,
\end{equation}
\end{subequations}
due to rotational invariance,
with a minus sign to anticipate the counteraction of the diffusive mass flux $\textbf{U}_i$ to the gradient of density $n_i$.

Thanks to the linearity of the collision integrals $I$, one gets
\begin{align*}
I&\left[\phi^T_i+\phi^S_i+\phi^k_i,~\phi^T_j+\phi^S_j+\phi^k_j\right] \\
&= I\left[\phi^T_i,\phi^T_j\right]+I\left[\phi^S_i,\phi^S_j\right]+I\left[\phi^k_i,\phi^k_j\right],
\end{align*}
and $\phi_i = \phi^T_i+\phi^S_i+\sum_k \phi^k_i$ is a solution when all the gradients are present. We shall take for granted that this solution exists and is unique provided that it satisfies the constraints Eq.\,\eqref{eq:ChapmanAnsatz}.

Consequently, the solution to the linear system of equations, Eq.\,\eqref{eq:eq_phi}--\eqref{eq:driving}, is therefore of the following form
\begin{align}
\label{eq:phi_grad}
\nonumber
\phi_i
= &- K_i^T(c) ~ \textbf{c} \cdot \nabla \log T \\
&- K_i^S(c) ~ \textbf{c} \otimes \textbf{c} : \rttensor S \\
&- \sum_k K^k_i(c) ~ \textbf{c} \cdot \nabla n_k.
\nonumber
\end{align}

The constraints of the Chapman-Enskog ansatz, Eq.\,\eqref{eq:ChapmanAnsatz}, translate into the following constraints on the functions $K^T$, $K^S$, and $K^k$
\begin{subequations}
\begin{align}
\int M_i(c) \, K_i^S(c) \,c^2\,d^3 \textbf{c} &=0,\\
\sum_i m_i\int  M_i(c) \, K^T_i(c)\, c^2\, d^3 \textbf{c} &=0,\\
\sum_i m_i\int  M_i(c) \, K^k_i(c)\, c^2\, d^3 \textbf{c} &=0,\\
\sum_i m_i \int  M_i(c) \, K_i^S(c) \,c^4\,d^3 \textbf{c} &=0.
\end{align}
\end{subequations}

Traditionally, the unknown functions $ K_i^T$, $K^S_i$, and $K^k_i$ are developed on a basis of Sonine polynomials, since at the time of these developments the digital computer was not discovered yet. We shall not dwell with the solution of this system, but we shall assume known the solution and derive its relationship with the closure of fluid equations. 

\section{Navier-Stokes equations}
\label{sec:TransportCoeff}

With the solution of the linearized Boltzmann equation expressed as a function of the gradients of densities $n_i$, fluid velocity $\textbf{u}$, and temperature $T$, Eq.\,\eqref{eq:phi_grad}, the closure relations of the fluid equations, Eq.\,\eqref{eq:MassFlux}, \eqref{eq:MomentumFlux}, and \eqref{eq:HeatFlux}, can be computed as functions of these gradients. The final results represent the \textit{constitutive relations} known as Fick's law, Fourier's law, and Newton's law, with their associated transport coefficients of diffusion, thermal conductivity, and viscosity, respectively. 

As a first step, it is useful to compute the rate of entropy production in order to identify the different dissipation mechanisms that involve the transport coefficients.

\subsection{H-theorem and rate of entropy production}
\label{sec:Entropy}

Boltzmann generalized the thermodynamic concept of entropy $S$ to non-equilibrium situations through the $H$ functional, defined by
\begin{equation}
H = - \dfrac{S}{k} = \sum_i \int f_i \log f_i \, d^3 \textbf{v}_i.
\end{equation}
The rate of entropy production $\Gamma_S$, in $k$ unit, is 
\begin{subequations}
\begin{equation}
\Gamma_S = - \dfrac{dH}{dt} = - \sum_i \int (\log f_i + 1) \, \dfrac{df_i}{dt} \, d^3 \textbf{v}_i,
\end{equation}
where the substantial time derivative $df_i/dt = \partial_t f_i  +  \textbf{v}_i\cdot \nabla f_i$ can be replaced by the collision integrals of Boltzmann equation to give
\begin{equation}
\Gamma_S = - \sum_{i,j} \int (\log f_i(\textbf{v}_{i}) + 1) \, J[f_i, f_j](\textbf{v}_i) \, d^3 \textbf{v}_i.
\end{equation}
The rate of entropy production $\Gamma_S$ involve a summation over every binary collisions of the moment of the collision integral with the function $K(\textbf{v}) = \log f_i(\textbf{v}) + 1$. As in Sec.\,\ref{sec:Invariants}, Eq.\,\eqref{eq:invariant2bis}, and App.\,\ref{app:moments}, it can be rewritten as
\begin{align}
\Gamma_S = \dfrac{1}{4} \sum_{i,j} \int (\log f_i' f_j' &- \log f_i f_j) (f_i'f'_j-f_if_j) \nonumber\\ 
&\times \mathbf{v}_{ij}\, 2 \pi \, b \, d b \, d^3 \textbf{v}_{j} \, d^3 \textbf{v}_i.
\end{align}
\end{subequations}
This expression is the proof of the H-theorem, related to the second principle of thermodynamics stating that the entropy is an increasing function of time, whatever the process (since the function $(x-y)(\log x - \log y)$ is always positive except for $x=y$ where it vanishes).

We now examine the rate of entropy production for the solution $\phi_i$ of the linearized Boltzmann equation, Eqs.\,\eqref{eq:CE_ansatz} and \eqref{eq:phi_grad}. At first order ($\varepsilon = 0)$, it vanishes since $\phi_i = M_i$ is the Maxwellian distribution function at thermodynamic equilibrium. At next order ($\varepsilon^2$), the collision operator $J$ is linearized to $I$, Eq.\,\eqref{eq:Boltzamnn_linear}, and the H-theorem reads
\begin{align}
\Gamma_S &= - \sum_{i,j} \int \phi_i(\textbf{v}_{i}) \, I[\phi_i, \phi_j](\textbf{v}_i) \, d^3 \textbf{v}_i\nonumber \\
&= \dfrac{1}{4} \sum_{i,j} \int M_i \,M_j\, (\phi_i'+ \phi_j' - \phi_i -\phi_j)  \\ 
&\hspace{25mm}\times (\phi_i' +\phi_j'-\phi_i-\phi_j) \nonumber\\ 
&\hspace{3cm}\times \mathbf{v}_{ij}\, 2 \pi \, b \, d b \, d^3 \textbf{v}_{j} \, d^3 \textbf{v}_i.\nonumber
\end{align}
The last expression is clearly positive, as expected. In App.\,\ref{app:LinColl}, a useful notation, called \textit{bracket integral}, was introduced for the summation over every binary collisions of the moments of the linearized collision integral 
\begin{subequations}
\begin{align}
\left[G \times F\right] &= - \sum_{i,j} \int G_i(\textbf{v}_{i}) \, I[F_i, F_j](\textbf{v}_i) \, d^3 \textbf{v}_i\\
&= \dfrac{1}{4} \sum_{i,j} \int M_i \,M_j\, (G_i'+ G_j' - G_i -G_j) \nonumber\\ 
&\hspace{25mm}\times (F_i' +F_j'-F_i-F_j) \nonumber\\ 
&\hspace{3cm}\times \mathbf{v}_{ij}\, 2 \pi \, b \, d b \, d^3 \textbf{v}_{j} \, d^3 \textbf{v}_i\nonumber\\
&= - \sum_{i,j} \int F_i(\textbf{v}_{i}) \, I[G_i, G_j](\textbf{v}_i) \, d^3 \textbf{v}_i\nonumber \\
&= \left[F \times G\right]\nonumber
\end{align}
This functional is also a bilinear form
\begin{equation}
\left[G \times (F_1+F_2)\right] = \left[G \times F_1\right]+\left[G \times F_2\right].
\end{equation}
\end{subequations}
Clearly, the rate of entropy production can be expressed shortly using this notation as
\begin{subequations}
\begin{equation}
\Gamma_S = \left[\phi \times \phi \right].
\end{equation}
In order to isolate the contributions of the different gradients of  densities $n_i$, fluid velocity $\textbf{u}$, and temperature $T$, we further substitute the expression of $\phi$ given by Eq.\,\eqref{eq:phi_grad}, under the shortcut form
\begin{equation}
\label{eq:phi_short}
\phi = \phi^T+\phi^S+\sum_k \phi^k.
\end{equation}
This leads to
\begin{align}
\Gamma_S &= \left[\phi^T \times \phi^T \right] + \left[\phi^S \times \phi^S \right]+ \sum_{k,l}\left[\phi^k \times \phi^l \right] \\
&+ 2\left( \left[\phi^T \times \phi^S \right] +  \sum_k\left[\phi^k \times \phi^T \right] +  \sum_k\left[\phi^k \times \phi^S \right]\right) \nonumber
\end{align}
The evaluation of each term is given in App.\,\ref{app:TransportCoeff} with the following results
\begin{equation}
 \left[\phi^T \times \phi^T \right] = \dfrac{1}{3}\, \left[\textbf{K}^T \cdot \textbf{K}^T\right]\, |\nabla \log T|^2,
\end{equation}
\begin{equation}
 \left[\phi^S \times \phi^S \right] = \dfrac{2}{15}\, \left[\rttensor{K}^S : \rttensor{K}^S\right]\, \rttensor S : \rttensor S,
\end{equation}
\begin{equation}
 \left[\phi^k \times \phi^l \right] = \dfrac{1}{3}\, \left[\textbf{K}^k \cdot \textbf{K}^l\right]\, \nabla n_k \cdot \nabla n_l,
\end{equation}
\begin{equation}
 \left[\phi^T \times \phi^S \right] = 0,
\end{equation}
\begin{equation}
 \left[\phi^k \times \phi^T \right] = \dfrac{1}{3}\, \left[\textbf{K}^k \cdot \textbf{K}^T\right]\, \nabla n_k \cdot \nabla \log T,
\end{equation}
\begin{equation}
 \left[\phi^k \times \phi^S \right] = 0,
\end{equation}
\end{subequations}
with obvious generalizations of the $\times$ notation in the bracket integrals to the scalar product for vectors, and the tensor reduction (see App.\,\ref{app:LinColl}). The coefficients of the different gradients are related to the thermal conductivity, the viscosity, the diffusion, and the thermal diffusion, respectively.

\subsection{Thermal and mass diffusions}

By introducing in the definition of the peculiar velocity $\textbf{U}_i$ of species $i$, Eq.\,\eqref{eq:MassFlux}, the expression of the distribution functions, Eqs.\,\eqref{eq:CE_ansatz} and \eqref{eq:phi_grad}, we identify the coefficients of mass diffusion $D_{ij}$ and thermal diffusion $D_{Ti}$
\begin{align}
\label{eq:Ui}
n_i \textbf{U}_i =  &\int \textbf{c} \,  M_i(c)\,\phi_i(\textbf{c}) \, d^3\textbf{c} \\
 = &- \int \textbf{c} \,  M_i(c)\,K_i^T(c) ~ \textbf{c} \cdot \nabla \log T\, d^3\textbf{c}
\nonumber \\
& - \int \textbf{c} \,  M_i(c)\,K_i^S(c) ~ \textbf{c} \otimes \textbf{c} : \rttensor S \, d^3\textbf{c}
\nonumber \\
& - \sum_j \int \textbf{c} \, M_i(c)\, K^j_i(c) ~ \textbf{c} \cdot \nabla n_j \, d^3\textbf{c}
\nonumber \\
 = & - n_i\,D_{Ti} \, \nabla \log T-\sum_j D_{ij} \, \nabla n_j,
\nonumber
\end{align} 
where the term involving $K_i^S$ vanishes as the integral of an odd function of $\textbf{c}$, and
\begin{align}
\label{eq:DTi}
n_i D_{Ti} &=- \dfrac{1}{3} \,  \int  M_i(c)~\textbf{K}_i^T(\textbf{c}) \cdot \textbf{c}\, d^3\textbf{c},\\
&= \dfrac{4 \pi}{3} \, \int_0^{\tiny\infty} M_i(c)\,K_i^T(c) \,\, c^4 \,  dc,\nonumber
\end{align}
\begin{align}
\label{eq:Dij}
D_{ij}  &= - \dfrac{1}{3} \, \int M_i(c)\,\textbf{K}_{i}^j(\textbf{c}) \cdot \textbf{c} \, d^3\textbf{c},\\
&= \dfrac{4 \pi}{3} \, \int_0^{\tiny\infty} M_i(c)\,K_{ij}^n(c) \,\, c^4 \,  dc,\nonumber
\end{align}
where Eq.\,\eqref{eq:IntegralVector} has been used.

These coefficients are not independent but respect the following \textit{sum rules}
\begin{subequations}
\label{eq:SumRule}
\begin{equation}
\sum_i m_i D_{ij} = 0,
\end{equation} 
\begin{equation}
\sum_i m_i n_i D_{Ti} = 0.
\end{equation}
\end{subequations} 
Indeed, by using $\rho \textbf{u} = \sum_i \rho_i \textbf{u}_i$ and $\textbf{U}_i = \textbf{u}_i - \textbf{u}$, to sum over the species the mass conservation equations of each species, Eq.\,\eqref{eq:mass_eq}
\begin{subequations}
\begin{equation}
\partial_t \rho_i + \nabla \cdot \left( \rho_i \textbf{u} \right) + \nabla \cdot \left( \rho_i \textbf{U}_i \right)  = 0,
\end{equation} 
one recovers the total mass conservation 
\begin{equation}
\label{symetry_relation}
\partial_t \rho + \nabla \cdot \left( \rho \textbf{u} \right) = 0.
\end{equation} 
This translates to the following constraint on the transport coefficients
\begin{align}
\label{eq:UiConstraint}
0 =  \sum_i & \rho_i \textbf{U}_i = \sum_i m_i n_i \textbf{U}_i  \\
& = - \sum_{ij} m_i D_{ij} \, \nabla n_j - \sum_i  m_i n_i D_{Ti} \, \nabla \log T,
\nonumber
\end{align} 
\end{subequations}
that should be verified whatever the values taken by the different gradients. 

The last step is to make contact with the rate of entropy production. To this end, we evaluate the following bracket integrals
\begin{align}
\left[\textbf{K}^l \cdot \textbf{K}^k\right] &= - \sum_{i,j} \int \textbf{K}^l_i(\textbf{c}_i) \cdot I[\textbf{K}^k_i, \textbf{K}^k_j](\textbf{c}_i) \, d^3 \textbf{c}_i\\
&=- \sum_{i} \int M_i\,\dfrac{1}{n_i}\, \left( \delta_{ik} - \dfrac{m_i n_i}{\rho}\right)\, \textbf{K}^l_i \cdot \textbf{c}_i \, d^3 \textbf{c}_i\nonumber\\
&= 3\, \dfrac{D_{kl}}{n_k} -  \dfrac{3}{\rho}\,\sum_i m_i D_{il} \nonumber \\
&= 3\, \dfrac{D_{kl}}{n_k} \nonumber
\end{align}
using Eq.\,\eqref{eq:Kij}, \eqref{eq:Dij}, and \eqref{eq:SumRule}. Due to the symmetry of the bracket integrals, this results in the symmetry property
\begin{equation}
\dfrac{D_{ij}}{n_i} =\dfrac{D_{ji}}{n_j} = \dfrac{1}{3} \, \left[\textbf{K}^i \cdot \textbf{K}^j\right].
\end{equation}

Likewise, we evaluate the bracket integrals
\begin{align}
\left[\textbf{K}^T \cdot \textbf{K}^k\right] &= - \sum_{i,j} \int \textbf{K}^T_i(\textbf{c}_i) \cdot I[\textbf{K}^k_i, \textbf{K}^k_j](\textbf{c}_i) \, d^3 \textbf{c}_i\\
&=- \sum_{i} \int M_i\,\dfrac{1}{n_i}\, \left( \delta_{ik} - \dfrac{m_i n_i}{\rho}\right)\, \textbf{K}^T_i \cdot \textbf{c}_i \, d^3 \textbf{c}_i\nonumber\\
&= 3\, D_{Tk} -  \dfrac{3}{\rho}\,\sum_i m_i n_i D_{Ti} \nonumber \\
&= 3\, D_{Tk} \nonumber
\end{align}
using Eq.\,\eqref{eq:Kij}, \eqref{eq:DTi}, and \eqref{eq:SumRule}. Due to the symmetry of the bracket integrals, we shall use the following property in the evaluation of the heat flux
\begin{equation}
\label{eq:DTiBracket}
D_{Ti} = \dfrac{1}{3} \, \left[\textbf{K}^T \cdot \textbf{K}^i\right] = \dfrac{1}{3} \, \left[\textbf{K}^i \cdot \textbf{K}^T\right].
\end{equation}

\subsection{Viscosity}

We proceed by introducing in the definition of the pressure tensor,  \eqref{eq:MomentumFlux},  the expression of the distribution functions, Eqs.\,\eqref{eq:CE_ansatz} and \eqref{eq:phi_grad}, to identify the coefficient of viscosity $\eta$
\begin{align}
\rttensor{\Pi} &- p\, \rttensor{Id} =  \sum_i m_i \int \textbf{c} \otimes \textbf{c} \, M_i(c)\,\phi_i(\textbf{c}) \, d^3 \textbf{c} \\
 = &- \sum_i m_i \int \textbf{c} \otimes \textbf{c}\,  M_i(c)\,K_i^T(c) ~ \textbf{c} \cdot \nabla \log T\, d^3\textbf{c}
\nonumber \\
& - \sum_i m_i \int \textbf{c} \otimes \textbf{c} \,  M_i(c)\,K_i^S(c)   ~ \textbf{c} \otimes \textbf{c} : \rttensor S \, d^3\textbf{c}
\nonumber \\
& - \sum_{ij} m_i \int \textbf{c} \otimes \textbf{c} \,  M_i(c)\,K^j_i(c) ~ \textbf{c} \cdot \nabla n_j \, d^3\textbf{c}
\nonumber \\
= & -2\,\eta \, \rttensor S,
\nonumber
\end{align}
where the terms involving $K_i^T$ and $K_i^j$ vanish as integrals of an odd function of $\textbf{c}$, and
\begin{align}
\eta  &= - \dfrac{1}{15} \sum_i m_i\,  \int M_i(c)\,\rttensor{K}_i^S(\textbf{c}) : \textbf{c} \otimes \textbf{c}  \, d^3\textbf{c}, \nonumber \\
&= \sum_i \dfrac{m_i}{2}\, \dfrac{8 \pi}{15} \int_0^{\tiny\infty} M_i(c)\,K_i^S(c) \,\, c^6 \, dc,
\end{align}
where Eq.\,\eqref{eq:IntegralTensor} has been used.

The connection with the rate of entropy production is made by evaluating the bracket integral
\begin{align}
&\left[\rttensor{K}^S : \rttensor{K}^S\right]= - \sum_{i,j} \int \rttensor{K}^S_{~i}(\textbf{c}_i) : I[\rttensor{K}^S_{~i}, \rttensor{K}^S_{~j}](\textbf{c}_i) \, d^3 \textbf{c}_i \nonumber \\
&=- \sum_{i} \int M_i\,\dfrac{m_i}{k T} ~ \rttensor{K}^S_{~i} : \textbf{c}_i \otimes \textbf{c}_i \, d^3 \textbf{c}_i\nonumber\\
&= \dfrac{15}{k T}\, \eta, 
\end{align}
using Eq.\,\eqref{eq:Kij}, \eqref{eq:DTi}, and \eqref{eq:SumRule}. This last expression warrants that the coefficient of viscosity is positive.
 
 \subsection{Thermal conductivity}

When introducing in the definition of the heat flux, Eq.\, \eqref{eq:HeatFlux}, the expression of the distribution functions, Eqs.\,\eqref{eq:CE_ansatz} and \eqref{eq:phi_grad}, there are different choices for the definition of the coefficient of thermal conductivity. We choose the one coming from the rate of entropy production, which is formally identical in both cases of the mixtures and the pure elements. The first step here consists in evaluating the bracket integral
\begin{align}
\left[\textbf{K}^T \cdot \textbf{K}^T\right] & =  - \sum_{i,j} \int \textbf{K}^T_i(\textbf{c}_i) \cdot I[\textbf{K}^T_i, \textbf{K}^T_j](\textbf{c}_i) \, d^3 \textbf{c}_i\\
=& - \sum_{i} \int M_i\,\left( \dfrac{m_i c_i^2}{2 k T}- \dfrac{3}{2}-\dfrac{m_i}{m}\right)\, \textbf{K}^T_i \cdot \textbf{c}_i \, d^3 \textbf{c}_i,\nonumber
\end{align}
and comparing it with the definition of the  heat flux
\begin{align}
\textbf{q}  = &\sum_i \int \dfrac{1}{2} m_i c^2\, \textbf{c} \,M_i(c)\, \phi_i(\textbf{c}) \, d^3 \textbf{c},\\
 = &- \sum_i \int \dfrac{1}{2} m_i c^2\,\textbf{c} \,  M_i(c)\,K_i^T(c) ~ \textbf{c} \cdot \nabla \log T\, d^3\textbf{c}
\nonumber \\
& - \sum_i \int \dfrac{1}{2} m_i c^2\,\textbf{c} \,  M_i(c)\,K_i^S(c) ~ \textbf{c} \otimes \textbf{c} : \rttensor S \, d^3\textbf{c}
\nonumber \\
& - \sum_{i,j} \int \dfrac{1}{2} m_i c^2\,\textbf{c} \,  M_i(c)\,K^j_i(c) ~ \textbf{c} \cdot \nabla n_j \, d^3\textbf{c}
\nonumber \\
= &\,\dfrac{3}{2} k T \sum_i n_i \textbf{U}_i - \lambda \, \nabla T - k T \, \sum_j  D_{jT} \, \nabla n_j.
\nonumber
\end{align}
where the term involving $K_i^S$ vanishes as the integral of an odd function of $\textbf{c}$, and
\begin{align}
\lambda  &= \dfrac{k}{3}\,\left[\textbf{K}^T \cdot \textbf{K}^T\right]\\
&= \dfrac{4 \pi k}{3} \sum_{i} \int_0^{\tiny\infty} M_i\,\left( \dfrac{m_i c^2}{2 k T}- \dfrac{3}{2}-\dfrac{m_i}{m}\right)\, K^T_i(c)\, c^4 \, dc, \nonumber
\end{align}
where Eq.\,\eqref{eq:IntegralVector} has been used. This last expression warrants that the coefficient of thermal conductivity is positive.

The remaining terms of the heat flux are
\begin{subequations}
\begin{align}
\textbf{q} &+ \lambda \, \nabla T = \\
&- \sum_i \int k T \left(\dfrac{3}{2}+\dfrac{m_i}{m}\right)\,\textbf{c} \,  M_i(c)\,K_i^T(c) ~ \textbf{c} \cdot \nabla \log T\, d^3\textbf{c} \nonumber \\
& - \sum_{i,j} \int \dfrac{1}{2} m_i c^2\,\textbf{c} \,  M_i(c)\,K^j_i(c) ~ \textbf{c} \cdot \nabla n_j \, d^3\textbf{c}, \nonumber
\end{align}
where we introduce the mass flux $\textbf{U}_i$ as expressed in Eq.\,\eqref{eq:Ui} and recalled here
\begin{align}
n_i \textbf{U}_i = &- \int \textbf{c} \,  M_i(c)\,K_i^T(c) ~ \textbf{c} \cdot \nabla \log T\, d^3\textbf{c}
\nonumber \\
& - \sum_j \int \textbf{c} \,  M_i(c)\,K^j_i(c) ~ \textbf{c} \cdot \nabla n_j \, d^3\textbf{c},
\nonumber
\end{align} 
leading to
\begin{align}
\textbf{q} &+ \lambda \, \nabla T = \dfrac{3}{2} k T \sum_i n_i \textbf{U}_i + \dfrac{k T}{m} \sum_i n_i m_i \textbf{U}_i  \\
&+ \dfrac{3}{2} k T \sum_{i,j} \int \textbf{c} \,  M_i(c)\,K^j_i(c) ~ \textbf{c} \cdot \nabla n_j \, d^3\textbf{c}\nonumber \\
&+ \dfrac{k T}{m} \sum_{i,j} m_i \int \textbf{c} \,  M_i(c)\,K^j_i(c) ~ \textbf{c} \cdot \nabla n_j \, d^3\textbf{c}\nonumber \\
& - \sum_{i,j} \int \dfrac{1}{2} m_i c^2\,\textbf{c} \,  M_i(c)\,K^j_i(c) ~ \textbf{c} \cdot \nabla n_j \, d^3\textbf{c}. \nonumber
\end{align}
Recall the constraint related to the total mass conservation, Eq.\,\eqref{eq:UiConstraint}, \textit{i.e.} $\sum_i m_i n_i \textbf{U}_i = 0$. Then, one is left with
\begin{align}
&\textbf{q} + \lambda \, \nabla T - \dfrac{3}{2} k T \sum_i n_i \textbf{U}_i   \\
&= - k T\sum_{i,j} \int \left(\dfrac{m_i c^2}{2 k T} - \dfrac{3}{2} - \dfrac{m_i}{m}\right)\,\nonumber \\
& \hspace{3cm}\times\textbf{c} \,  M_i(c)\,K^j_i(c) ~ \textbf{c} \cdot \nabla n_j \, d^3\textbf{c} \nonumber \\
&= - \dfrac{k T}{3} \sum_j \left[\textbf{K}^j \cdot \textbf{K}^T\right] \nabla n_j = - k T \sum_j D_{Tj} \nabla n_j, \nonumber
\end{align}
\end{subequations}
using Eq.\,\eqref{eq:IntegralVector} and the symmetry of the bracket integrals, Eq.\,\eqref{eq:DTiBracket}.

\section{Emergence of phenomena in multicomponent mixtures}
\label{sec:OneTwoThree}

We shall see, in this section, how new dissipative phenomena appear as the number of components in a mixture increases from the case of pure elements, to binary mixtures, and beyond three species. For a fluid made of only one component,
there are only the transport coefficients of viscosity $\eta$ and thermal conductivity $\lambda$. For a mixture, there appear additional transport coefficients of thermal diffusion $D_{Ti}$ and mutual diffusion $D_{ij}$. For mixtures with more than two components, the interdiffusion coefficients $D_{ij}$ exhibit a complex behavior that goes beyond \textit{Fick's law}. 

\subsection{Pure elements}

For pure elements, the Boltzmann equation involves only one distribution function $f(t, \textbf{r}, \textbf{v})$. The peculiar velocity $\textbf{u}_i$ coincides with the fluid velocity $\textbf{u}$, $\textbf{U}_i = \textbf{u}_i - \textbf{u} = 0$. The driving force, in the Chapman-Enskog formulation, no longer involves the density gradient
\begin{align}
\dfrac{1}{M}\,(\partial_t M + \textbf{v} \cdot \nabla M)
=& \left( \dfrac{m \textbf{c}^2}{2 k T}- \dfrac{5}{2}\right) \, \textbf{c} \cdot \nabla \log T \nonumber\\
&+ \dfrac{m}{k T} ~ \textbf{c} \otimes \textbf{c} : \rttensor{S},
\end{align} 
and the mass conservation equation reduces to the total mass conservation equation. This equation is also valid for mixtures
\begin{equation}
\partial_t \rho + \nabla \cdot \left( \rho \textbf{u} \right) = 0.
\end{equation}

The momentum conservation equation is already of a form valid for mixtures
\begin{subequations}
\begin{equation}
\partial_t (\rho \textbf{u}) + \nabla\cdot\left( \rho \textbf{u} \otimes \textbf{u} + \rttensor{\Pi} \right)=0,
\end{equation}
with \textit{Newton's constitutive relation} for the pressure tensor
\begin{equation}
\rttensor{\Pi} =  p\, \rttensor{Id}  -2\,\eta \, \rttensor S.
\end{equation}
\end{subequations}

The energy conservation equation is the one most affected by the emergence of new dissipative phenomena. This energy dissipation emerges through the heat flux $\textbf{q}$. For pure elements, this flux is given by \textit{Fourier's constitutive relation}
\begin{subequations}
\begin{equation}
\textbf{q}_\text{pure} = - \lambda \nabla T.
\end{equation}
To highlight the emergence of the dissipative phenomena in a mixture, we explicit the heat flux in the energy conservation equation, which is given here for pure elements
\begin{equation}
\partial_t E + \nabla \cdot \left( E \,\textbf{u} + \rttensor{\Pi} \cdot \textbf{u} - \lambda \nabla T \right) = 0.
\end{equation}
\end{subequations}

\subsection{Binary mixtures}

In the case of the binary mixtures, the total mass conservation equation can be replaced by the species mass conservation equations
\begin{subequations}
\begin{align}
&\partial_t \rho_1 + \nabla \cdot \left( \rho_1 \textbf{u} \right) + \nabla \cdot \left( \rho_1 \textbf{U}_1 \right)  = 0,\nonumber\\
&\partial_t \rho_2 + \nabla \cdot \left( \rho_2 \textbf{u} \right) + \nabla \cdot \left( \rho_2 \textbf{U}_2 \right)  = 0,
\end{align} 
with the closure relations
\begin{equation}
\label{closure_binary_eq1}
\begin{aligned}
&\rho_1 \textbf{U}_1 = -m_1 D_{11} \, \nabla n_1 -m_1 D_{12} \, \nabla n_2 - \rho_1\,D_{1T} \, \nabla \log T,\\
&\rho_2 \textbf{U}_2 = - m_2 D_{21} \, \nabla n_1 - m_2 D_{22} \, \nabla n_2- \rho_2\,D_{2T} \, \nabla \log T,
\end{aligned}
\end{equation}
However, all the coefficients of interdiffusion $D_{ij}$ and thermal diffusion $D_{Ti}$ are not independent. 
They obey the sum rules, Eq.\, \eqref{eq:SumRule}, which read
\begin{equation}
\begin{aligned}
&m_1 D_{11} + m_2 D_{21}  = 0, \\
&m_1 D_{12} + m_2 D_{22}  = 0, \\
&\rho_1 D_{1T} = - \rho_2 D_{2T} = \rho \, D_T,
\end{aligned}
\end{equation}
where we have defined a unique coefficient of thermal diffusion, $D_T$. They also exhibit symmetry properties, Eq.\,  \eqref{symetry_relation}. In particular, one gets 
\begin{equation}
  D_{12}/n_1 = D_{21}/n_2 = - D /n,
\end{equation}
where we have defined a unique coefficient of interdiffusion, $D$. With these unique coefficients, the closure relations read
\begin{equation}
\begin{aligned}
&\rho_1 \textbf{U}_1 = -\rho_2 \, D \, \dfrac{\nabla n_1}{n} + \rho_1 \, D \, \dfrac{\nabla n_2}{n} - \rho\,D_{T} \, \nabla \log T,\\
&\rho_2 \textbf{U}_2 = + \rho_2 \, D \, \dfrac{\nabla n_1}{n} - \rho_1 \, D \, \dfrac{\nabla n_2}{n} + \rho\,D_{T} \, \nabla \log T.
\end{aligned}
\end{equation}
As expected, one recovers
\begin{equation}
\rho_2 \textbf{U}_2 = - \rho_1 \textbf{U}_1 = -\rho \textbf{U},
\end{equation}
defining the peculiar velocity $\textbf{U}$.
\textit{Fick's law} appears when expressing the species densities $n_i$ in term of the concentrations $x_i = n_i/n$ since in the binary mixture $x_2 = 1 - x_1$ and $\nabla x_2 = - \nabla x_1$
\begin{equation}
\begin{aligned}
&\rho_1 \textbf{U}_1 = -\rho \, D \, \nabla x_1 - (x_1\rho_2 - x_2 \rho_1) \, D \, \dfrac{\nabla n}{n} - \rho\,D_{T} \, \nabla \log T,\\
&\rho_2 \textbf{U}_2 = -\rho \, D \, \nabla x_2 + (x_1\rho_2 - x_2 \rho_1) \, D \, \dfrac{\nabla n}{n} + \rho\,D_{T} \, \nabla \log T.
\end{aligned}
\end{equation}
In this last equation, the terms proportional to the gradient of total density lead to the barodiffusion when the equation of state is used to relate the total density to the pressure and the temperature. In the ideal gas case, already used to derived the driving forces in App.\,\ref{app:Snider}, one gets
\begin{equation*}
\dfrac{\nabla n}{n} = \nabla \log n = \nabla \log P - \nabla \log T.
\end{equation*}
\end{subequations}
This gives rise to an additional contribution to the thermal diffusion.

The energy conservation equation is modified with respect to the case of pure elements by the additional sources of dissipation appearing in the heat flux
\begin{subequations}
\begin{align}
\textbf{q} = &- \lambda \, \nabla T \\
&+ \dfrac{3}{2} k T \left(\dfrac{1}{m_1}-\dfrac{1}{m_2}\right)\rho \textbf{U} \nonumber\\
&- k T \, \rho D_T \left(\dfrac{1}{m_1} \, \dfrac{\nabla n_1}{n_1} - \dfrac{1}{m_2} \, \dfrac{\nabla n_2}{n_2}\right).\nonumber
\end{align}
This can be interpreted as a failure of \textit{Fourier's law}. These contributions to the heat flux should not be omitted to prevent a loss of energy conservation. For binary mixtures, the energy conservation equation reads
\vskip 1cm
\begin{widetext}
\begin{equation}
\partial_t E + \nabla \cdot \left( E \,\textbf{u} + \rttensor{\Pi} \cdot \textbf{u} - \lambda \nabla T + \dfrac{3}{2} k T \left(\dfrac{1}{m_1}-\dfrac{1}{m_2}\right)\rho \textbf{U}- k T \, \rho D_T \left(\dfrac{1}{m_1} \, \dfrac{\nabla n_1}{n_1} - \dfrac{1}{m_2} \, \dfrac{\nabla n_2}{n_2}\right)\right) = 0.
\end{equation}
\end{widetext}
\end{subequations}

\subsection{Ternary mixtures}

For more than two species in a mixture, the formulation does not change so much, but \textit{Fick's constitutive relation} must be generalized. It fails to predict some new diffusion phenomena. To illustrate this breakdown, we consider the equations of mass conservation for each species of a ternary mixture without the presence of gradients of temperature and total density. This alleviates the exposition. With these restrictions,  the species mass conservation equations reduce to
\begin{subequations}
\begin{equation}
\begin{aligned}
&\partial_t \rho_1 + \nabla \cdot \left( \rho_1 \textbf{u} \right) + \nabla \cdot \left( \rho_1 \textbf{U}_1 \right)  = 0,\\
&\partial_t \rho_2 + \nabla \cdot \left( \rho_2 \textbf{u} \right) + \nabla \cdot \left( \rho_2 \textbf{U}_2 \right)  = 0,\\
&\partial_t \rho_3 + \nabla \cdot \left( \rho_3 \textbf{u} \right) + \nabla \cdot \left( \rho_3 \textbf{U}_3 \right)  = 0,
\end{aligned}
\end{equation}
with
\begin{equation}
\begin{aligned}
&\rho_1 \textbf{U}_1 = -m_1 D_{11} \, \nabla n_1 -m_1 D_{12} \, \nabla n_2 -m_1 D_{13} \, \nabla n_3,\\
&\rho_2 \textbf{U}_2 = - m_2 D_{21} \, \nabla n_1 - m_2 D_{22} \, \nabla n_2 - m_2 D_{23} \, \nabla n_3,\\
&\rho_3 \textbf{U}_3 = - m_3 D_{31} \, \nabla n_1 - m_3 D_{32} \, \nabla n_2 - m_3 D_{33} \, \nabla n_3,
\end{aligned}
\end{equation}

The symmetry of the interdiffusion coefficients and their sum rules write
\begin{equation}
\begin{aligned}
& D_{12}/n_1 = D_{21}/n_2 = - D_{(12)}/n, \\
& D_{13}/n_1 = D_{31}/n_3 = - D_{(13)}/n,\\
& D_{23}/n_2 = D_{32}/n_3 = - D_{(23)}/n,
\end{aligned} 
\end{equation}
and
\begin{equation}
\begin{aligned}
&m_1\, D_{11} = - m_2\, D_{21} - m_3\, D_{31},\\
&m_2\, D_{22} = -m_1\, D_{12} - m_3\, D_{32},\\
&m_3\, D_{33} = - m_1\, D_{13} - m_2\, D_{23}.
\end{aligned} 
\end{equation}
Altogether, only three coefficients of mutual diffusion are independent, $D_{(12)}$, $D_{(13)}$, and $D_{(23)}$. Therefore, the closure relations reduce to 
\vskip 1cm
\begin{widetext}
\begin{equation}
\begin{aligned}
\rho_1 \textbf{U}_1 = & - \left[ \rho_2 D_{(12)} + \rho_3 D_{(13)} \right] \,  \nabla x_1 + \rho_1 D_{(12)} \, \nabla x_2 + \rho_1 D_{(13)} \, \nabla x_3, \\
\rho_2 \textbf{U}_2 = & - \left[ \rho_1 D_{(12)}  + \rho_3 D_{(23)} \right] \, \nabla x_2  + \rho_2 D_{(12)} \, \nabla x_1 
 + \rho_2 D_{(23)} \, \nabla x_3, \\
\rho_3 \textbf{U}_3 = & - \left[ \rho_1 D_{(13)}  + \rho_2 D_{(23)} \right] \, \nabla x_3  + \rho_3 D_{(13)} \, \nabla x_1 + \rho_3 D_{(23)} \, \nabla x_2, 
\end{aligned}
\end{equation}
\end{widetext}
\end{subequations}
where additional terms appear besides the Fickian diffusion ($\textbf{U}_i$ proportional to $\nabla x_i$), leading to osmotic diffusion, reverse diffusion, or diffusion barrier \cite{Krishna2019}. For instance, there are circumstances where $\textbf{U}_i$ does not vanish even though $\nabla x_i$ vanishes.

\section{Conclusion}

The route, traced by Chapman and Enskog \cite{Chapman1970}, from the description of the binary collisions and the Boltzmann kinetic equations to the Navier-Stokes equations, is a long one. However, it is very instructive to follow it in order to highlight the most important physical arguments and assumptions.

The principal pillar of this edifice is the \emph{Knudsen number}, $\varepsilon$. Its very definition is possible when kinetic theory offers us the opportunity to define microscopic scales of time and space: the collision time $\tau$ and the mean free path $\lambda$. 

The criterion for reaching thermodynamic equilibrium can then be formulated: one can consider in equilibrium a volume, which is homogeneous over distances much larger than $\lambda$, and which is left out any solicitations for a time much greater than $\tau$. This criterion applies equally well out of the validity domain of kinetic theory, at least in order of magnitude.

With the \emph{Knudsen number}, $\varepsilon$,  the assumption of small gradients of the flow variables, that is invoked in the derivation from the Thermodynamics of irreversible processes, becomes an operational criterion, since it can be checked that the macroscopic scales of time and length, $T_0$ and $L_0$, associated with these gradients, are indeed much larger than $\tau$ and $\lambda$. The Euler equations can be used to estimate these gradients, for this comparison.

If $\varepsilon$ is small, and the conditions of applicability of kinetic theory are met, the Boltzmann equations can be linearized with  source terms, known as driving forces, arising from the decoupling of orders in $\varepsilon$. These driving forces dictate the form of the general solution, and of the transport coefficients.

However, the edifice is very fragile. One often takes for grant that the  \emph{Knudsen number} is small without further verification. The difficulties can arise when the thermodynamic equilibrium is not complete. 

In ICF, the ions and the electrons of the plasma must often be considered at different temperatures, for instance. This assumes that both distribution functions of ions and electrons relax to two Maxwellians at different temperatures, $T_i$ and $T_e$, quicker than the time required for the relaxation between $T_i$ and $T_e$. The fluid equations must therefore includes two equations for each energy of the ions and the electrons. As a result, additional terms appear coming from the collision integrals evaluated with Maxwellians at $T_i$ and $T_e$. These terms warrant that the system shall relax to a common temperature. In most situations, the resulting Euler equations are then dominated by the time scale $\tau_{ie}$ of temperature relaxation. Due to the large mass ratio between ions and electrons, it is however possible to define a small \emph{Knudsen number} in order to get linearized Boltzmann equations with driving forces as the source terms \cite{Simakov2016}.

In other circumstances, there is a decoupling of velocity. For instance, when two fluids meet at different velocities. This has an impact on the equations of momentum and energy conservation, with additional contributions coming from the collision integrals evaluated with Maxwellians centered at different velocities. The issue of defining a \emph{Knudsen number} in this case is much more involved. Different approaches are developed following either the thermodynamic route \cite{Bothe2015} or the kinetic one \cite{Larroche2021a} and confronting the hydrodynamic approximation to more detailed descriptions requiring microscopic simulations.

\section*{Data Availability Statement}

Data sharing is not applicable to this article as no new data were created or analyzed in this study.

\appendix

\section{Binary collisions}
\label{app:Collision}

Assume each collision between two particles of masses $m_i$ and $m_j$ can be assigned initial velocities $\textbf{v}_{i}$ and $\textbf{v}_{j}$, and final velocities $\textbf{v}'_{i}$ and $\textbf{v}'_{j}$. Conservation equations of  momentum and energy then read
\begin{align}
\label{eq:binary_momentum_energy}
m_i \textbf{v}_{i} + m_j \textbf{v}_{j} & = m_i \textbf{v}'_{i} + m_j \textbf{v}'_{j}, \\
\dfrac{1}{2} m_i \textbf{v}_{i}^2 + \dfrac{1}{2} m_j \textbf{v}_{j}^2 & = \dfrac{1}{2} m_i (\textbf{v}'_{i})^2 + \dfrac{1}{2} m_j (\textbf{v}'_{j})^2,
\nonumber
\end{align}
At this point, system \eqref{eq:binary_momentum_energy} represents four equations whereas there are six unknowns $\textbf{v}'_{i}$ and $\textbf{v}'_{j}$. Some additional information is then required. This information concerns what kind of interaction exists between the particles.

Consider that the two particles interact via a central force field $\textbf{F}(r)$, deriving from a potential $E_P(r)$, so that
\begin{equation}
 \textbf{F}(r) = - \nabla E_P.   
\end{equation}
Their motion follows Newton's law
\begin{align}
m_i \dfrac{d^2\textbf{r}_{i}}{dt^2} &= \textbf{F}(|\textbf{r}_{i} - \textbf{r}_{j}|),\\ \nonumber
m_j \dfrac{d^2\textbf{r}_{j}}{dt^2} &= -\textbf{F}(|\textbf{r}_{i} - \textbf{r}_{j}|).
\end{align}
The trajectory of both particles during the collision is easily derived in the center-of-mass (COM) frame, defined by COM ($\textbf{R}$, $\textbf{V}$) and relative ($\textbf{r}$, $\textbf{v}$) coordinates
\begin{eqnarray}\label{com}
&\textbf{R} = \dfrac{m_i \textbf{r}_{i}+m_j \textbf{r}_{j}}{m_i+m_j}, \quad \textbf{V} = \dfrac{m_i \textbf{v}_{i}+m_j \textbf{v}_{j}}{m_i+m_j},\\\nonumber
& \textbf{r} = \textbf{r}_{i} - \textbf{r}_{j}, \quad \quad \quad \,\,\, \textbf{v} = \textbf{v}_{i} - \textbf{v}_{j}.
\end{eqnarray}
The Jacobian of the transformation from the laboratory to the COM frames is equal to 1, so that in particular $d^3\textbf{v}_{i} \, d^3\textbf{v}_{j} = d^3\textbf{v} \, d^3\textbf{V}.$
In the COM frame, the equations of motion involve the reduced mass $m_{ij} = \dfrac{m_i m_j}{m_i+m_j}$  and read
\begin{subequations}
\begin{equation}\label{Newton}
m_{ij} \dfrac{d^2\textbf{r}}{dt^2} = m_{ij} \dfrac{d\textbf{v}}{dt} = \textbf{F}(r), 
\end{equation} 
for the relative motion, and 
\begin{equation}
\dfrac{d^2\textbf{R}}{dt^2} = \dfrac{d\textbf{V}}{dt} = 0,
\end{equation} 
\end{subequations}
for the COM motion. The latter equation expresses the \emph{momentum conservation}.

\begin{figure}[t!]
\begin{center}
\includegraphics[scale=0.45]{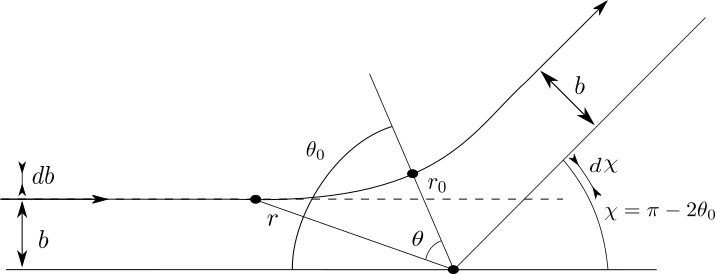}
\end{center}
\caption{Trajectory of a binary collision in the center of mass frame.}
\label{collision}
\end{figure}

\noindent The vector product with $\textbf{r}$ of Eq.\,\eqref{Newton} allows one to express the \emph{angular momentum conservation} and implies that the trajectory lies in a plane perpendicular to the angular momentum $(m_{ij} \textbf{r} \times \textbf{v})$
\begin{subequations}
\begin{equation}
\textbf{r} \times m_{ij} \dfrac{d^2\textbf{r}}{dt^2} =\dfrac{d}{dt}\left(m_{ij} \textbf{r} \times \dfrac{d\textbf{r}}{dt}\right) = \textbf{r} \times \textbf{F}(r) = 0.
\end{equation}
In the polar coordinates $(r, \theta)$ depicted in Fig.\,\ref{collision}, the position of the relative motion must verify the following constraint
\begin{equation}\label{angular_momentum}
m_{ij} \, r^2 \, \dfrac{d\theta}{dt} = m_{ij} \, b \, v_\infty,
\end{equation}
\end{subequations}
where $b$ is the impact parameter, the minimal distance between the particles if they do not interact and move in straight lines, and $v_\infty$ is the relative velocity before collision. Indeed, at large distance before collision, the relative position $\textbf{r} = x \, \textbf{e}_x + b \, \textbf{e}_y$, and the relative velocity $\textbf{v}_{ij} = v_\infty \textbf{e}_x$, for a cartesian system where the $x$-axis is along the initial relative velocity, and $\textbf{e}_y = \textbf{e}_z \times \textbf{e}_x$ with $\textbf{e}_z = (\textbf{v}_i \times \textbf{v}_j)/ v_i v_j $, a unit vector perpendicular to the trajectory plane.  

The scalar product with $\textbf{v} = d\textbf{r}/dt$ of Eq.\,\eqref{Newton} allows one to express the \emph{energy conservation}
\begin{subequations}
\begin{equation}
\dfrac{d\textbf{r}}{dt}\cdot m_{ij} \dfrac{d^2\textbf{r}}{dt^2} = \dfrac{d}{dt}\left(\dfrac{1}{2} m_{ij} \dfrac{d\textbf{r}}{dt}\cdot\dfrac{d\textbf{r}}{dt}\right) = \dfrac{d\textbf{r}}{dt}\cdot\textbf{F}(r),
\end{equation}  
with
\begin{equation}
\dfrac{d\textbf{r}}{dt}\cdot\textbf{F}(r) = -\dfrac{d\textbf{r}}{dt}\cdot\nabla E_P = -\dfrac{d E_P}{dt}.
\end{equation}  
This translates in polar coordinates to the second constraint
\begin{equation}
\label{energyCollision}
 \dfrac{1}{2}m_{ij}\left[\left(\dfrac{dr}{dt}\right)^2+r^2 \left(\dfrac{d\theta}{dt}\right)^2\right]+ E_P(r)= \dfrac{1}{2} m_{ij} v_\infty^2.
\end{equation}
\end{subequations}

A first consequence of the latter equation is that the relative velocity is conserved in modulus, equal to $v_\infty$. The only change produced by the elastic collision is to rotate this vector by an angle $\chi$, the scattering angle (Fig.\,\ref{collision}). In the Cartesian frame, the relative velocity before collision is 
\begin{subequations}
\begin{equation}
    \textbf{v}_{ij} = v_\infty \, \textbf{e}_x,
\end{equation}
and it is 
\begin{equation}
    \textbf{v}'_{ij} = v_\infty \left[ \cos(\chi) \, \textbf{e}_x + \sin(\chi)\,\textbf{e}_y \right],
\end{equation}
\end{subequations}
after collision.  

The velocities in the laboratory frame are given inverting Eq.\,\eqref{com}. Before collision, one can write the following relations
\begin{subequations}
\begin{align}
\textbf{v}_i = \dfrac{m_j}{m_i+m_j} \textbf{v}_{ij} + \textbf{V}, \hspace{0.3cm} \textbf{v}_j = -\dfrac{m_i}{m_i+m_j} \textbf{v}_{ij} + \textbf{V},
\end{align} 
 while after collision, 
\begin{align}
\label{FinalV}
\textbf{v}_i' = \dfrac{m_j}{m_i+m_j} \textbf{v}'_{ij} + \textbf{V}, \hspace{0.3cm} \textbf{v}_j' = -\dfrac{m_i}{m_i+m_j} \textbf{v}'_{ij} + \textbf{V}.
\end{align} 
\end{subequations}

We point out here that a relation between the scattering angle $\chi$ and the impact parameter $b$ can be obtained once the trajectory is parameterized through the relation between the radius $r$ and the angle $\theta$ (or $\chi$). 
 Indeed, using Eqs.\,\eqref{angular_momentum} and \eqref{energyCollision} to express separately the derivatives with time of the relative position $r$ and of the angle $\theta$, one gets
\begin{subequations}
\label{theta_r_t}
\begin{equation}
\dfrac{d\theta}{dt} = \dfrac{b \, v_\infty}{r^2}
\end{equation}

\begin{equation}
\left(\dfrac{dr}{dt}\right)^2 = v_\infty^2 \left( 1 - \dfrac{b^2}{r^2} - W(r)\right),
\end{equation}  
with
\begin{equation}
 W(r)= \dfrac{E_P(r)}{\dfrac{1}{2} m_{ij} v_\infty^2}.
\end{equation}  
\end{subequations}

We then define the turning point $r_0$ that cancels $dr/ dt$. It is the point of closest approach along the trajectory, solution to
\begin{equation}
    1 - \dfrac{b^2}{r_0^2} - W(r_0) = 0.
\end{equation}

It is convenient to define the origin of time $t$ and angle $\theta$ at the turning point. Indeed, the trajectory is symmetric with respect to this point, with a value of the angle $\theta =  -\, \theta_0$ before collision, when $t \to - \infty$ and $\theta =  + \theta_0$ after collision, when $t \to + \infty$. The deflection angle $\chi$ is then given by
\begin{subequations}
\label{relation_b_chi}
\begin{equation}
    \chi = \chi(b,v_\infty) = \pi - 2 \theta_0.
\end{equation}

To compute the trajectory from $t = 0$ to $+\infty$, an equation can be derived between the variations of the radius $r$ and the angle $\theta$ using Eq.\,\eqref{theta_r_t}

\begin{equation}
 \dfrac{dr}{d\theta}= \dfrac{r^2 }{b} \sqrt{1 - \dfrac{b^2}{r^2} - W(r)}.
\end{equation}  
Integrating the inverse relation giving $d\theta/dr$ from the turning point $r_0$ to $\infty$ gives the angle $\theta_0$ directly related to the deflection angle $\chi$
\begin{equation}
\theta_0 = \int_{r_0}^\infty \dfrac{1}{r^2 }\dfrac{b dr}{\sqrt{1 - \dfrac{b^2}{r^2} - W(r)}}.
\end{equation} 
\end{subequations}
As expected, this relation depends on the type of interaction between the particles, through the potential energy $E_P(r)$. It depends also on the relative velocity of the particles $v_\infty$.

When introducing Boltzmann's equation, we shall use Liouville's law, which expresses that the Jacobian of the transformation from the laboratory to the COM frames is equal to 1, in particular for velocities before and after collisions
\begin{equation}
\label{Liouville}
d^3\textbf{v}_{i} \, d^3\textbf{v}_{j} = d^3\textbf{v}_{ij} \, d^3\textbf{V} = d^3\textbf{v}'_{ij} \, d^3\textbf{V} = d^3\textbf{v}_{i}' \, d^3\textbf{v}_{j}'.
\end{equation}

\section{Boltzmann collision integral}
\label{app:Boltzmann}

\subsection{Empirical derivation}
\label{app:CollisionOperator}

Boltzmann gave an empirical derivation of his kinetic equation by considering the net effect between the collisions depleting the distribution function at a given velocity and the inverse collisions replenishing the distribution. 

In a fluid, being a dilute gas or a plasma, many binary collisions have to be considered. In App.\,\ref{app:Collision}, it is shown that a binary collision can be defined by the initial velocities $\textbf{v}_{i}$ and $\textbf{v}_{j}$ of the two colliding particles of species $i$ and $j$ and by their impact parameter $b$. Eq.\,\eqref{FinalV} gives the final velocities $\textbf{v}'_{i}$  and $\textbf{v}'_{j}$ in this collision. We shall compute the rate of these collisions, $\Gamma_{ij}\,d^3 \textbf{v}_{i}\,d^3 \textbf{v}_{j}\,db$ in the fluid.

It is helpful to consider each particles as a target for the collisions with the other particles, considered as projectiles. These projectiles are characterized by the velocity distribution function $f_i = f(t, \textbf{r}, \textbf{v}_{i})$, such that $(f_i d^3\textbf{v}_{i})$ gives the number of particles of mass $m_i$ with velocity between $\textbf{v}_{i}$ and $\textbf{v}_{i}+d^3\textbf{v}_{i}$, per unit volume, at time $t$ and position $\textbf{r}$. Now consider a flux of these particles, $\textbf{F}_{ij}$, incident onto a single scattering target of velocity $\textbf{v}_{j}$, and mass $m_j$, located at $\textbf{r}$ at time $t$, it is given by \begin{subequations}
\begin{equation}
\textbf{F}_{ij} = (f_i d^3\textbf{v}_{i})\, \textbf{v}_{ij}.
\end{equation}
Fig.\,\ref{collision} illustrates, in the COM frame, that the number $\dot N(\chi)\, d\chi$ of particles scattered per unit time between $\chi$ and $\chi + d\chi$, i.e. of velocities $\textbf{v}'_{i}$, is equal to the number of incident particles per unit time between $b$ and $b + db$, with $b$ related to $\chi$ by Eq.\,\eqref{relation_b_chi}, that is
\begin{equation}
\dot N(\chi)\, d\chi = 2 \pi \, b \,db \, F_{ij},
\end{equation}
The rate of collisions, $\Gamma_{ij}\,d^3 \textbf{v}_{i}\,d^3 \textbf{v}_{j}\,db$, must also account for the number $(f_j d^3\textbf{v}_{j})$ of scattering targets of mass $m_j$ and velocity between $\textbf{v}_{j}$ and $\textbf{v}_{j}+d^3\textbf{v}_{j}$, located at $\textbf{r}$ at time $t$
\begin{equation}
 \Gamma_{ij}\,d^3 \textbf{v}_{i}\,d^3 \textbf{v}_{j}\,db = (f_j d^3 \textbf{v}_{j}) \, \dot N(\chi)\, d\chi,
\end{equation}
So that finally one gets
\begin{equation}
 \Gamma_{ij}\,d^3\textbf{v}_{i}\,d^3 \textbf{v}_{j}\,db = (f_i d^3 \textbf{v}_{i})  (f_j d^3 \textbf{v}_{j}) \, v_{ij} \, 2 \pi \, b \, d b.
\end{equation}
\end{subequations}

Consider now the collisions with initial velocities $\textbf{v}'_{i}$ and $\textbf{v}'_{j}$ that restore, as final velocities, $\textbf{v}_{i}$ and $\textbf{v}_{j}$, with a rate equals to
\begin{equation}
 \Gamma'_{ij}\,d^3\textbf{v}'_{i}\,d^3 \textbf{v}'_{j}\,db = (f'_i d^3 \textbf{v}'_{i})  (f'_j d^3 \textbf{v}'_{j}) \, {v}_{ij} \, 2 \pi \, b \, d b,
\end{equation}
where $f'_i = f(t, \textbf{r}, \textbf{v}'_{i})$, and we use the fact that $\mathbf{v}'_{ij} = \mathbf{v}_{ij}$. Moreover, for elastic collisions, Liouville's law, Eq.\,\eqref{Liouville}, states that
\begin{equation}
d^3\textbf{v}'_{i}\,d^3\textbf{v}'_{j} = d^3\textbf{v}_{i}\,d^3\textbf{v}_{j}.
\end{equation}
The two rates are therefore equal at equilibrium, when the velocity distribution functions are Maxwellian, since then $f'_i f'_j = f_i f_j$ due to the conservation of energy in each binary collision, Eq.\,\eqref{eq:binary_momentum_energy}.
Out of equilibrium, the net rate of production of particles with velocity $\textbf{v}_{i}$ (from all the collisions with species $j$) per unit of phase space is the Boltzmann collision integral
\begin{align}
J[f_i, f_j](\textbf{v}_{i}) &= \int \left(\Gamma'_{ij} - \Gamma_{ij}\right)\, d^3 \textbf{v}_{j}\, d b, \\ \nonumber
 &= \int (f_i'f'_j-f_if_j) \, v_{ij}\, 2 \pi \, b \, d b \, d^3 \textbf{v}_{j}.
\end{align}

\subsection{Moments}
\label{app:moments}

Consider an arbitrary function $K(\textbf{v})$ and the integral 
\begin{align}
J_i[K, f_i, f_j] & = \int  K(\textbf{v}_{i}) \, J[f_i,f_j](\textbf{v}_{i}) \, d^3 \textbf{v}_{i} ,\\
& = \int K_i (f_i'f'_j-f_if_j) \,v_{ij}\,2 \pi b\,db\, d^3 \textbf{v}_{i} d^3 \textbf{v}_{j}. \nonumber
\end{align}
Now by changing variables from initial to final velocities, one gets
\begin{align}
\label{eq:invariant1}
J_i&[K, f_i, f_j] = \int K_i' (f_i f_j-f_i' f_j') \,v_{ij}'\,2 \pi b\,db\, d^3 \textbf{v}_{i}' d^3 \textbf{v}_{j}',\nonumber \\
&= \int K_i' (f_i f_j-f_i' f_j') \,v_{ij}\,2 \pi b\,db\, d^3 \textbf{v}_{i} d^3 \textbf{v}_{j},\\
&=\dfrac{1}{2} \int (K_i -K_i') \, (f_i' f_j'-f_i f_j) \,v_{ij}\,2 \pi b\,db\, d^3 \textbf{v}_{i} d^3 \textbf{v}_{j},\nonumber
\end{align}
due to Liouville's law for elastic collisions, Eq.\,\eqref{Liouville}. 

When this is summed over species $i$ and $j$, the indexes become dummy and the following relations hold
\begin{align}
\label{eq:invariant2}
J[K] &= \sum_{i, j} J_i[K, f_i, f_j]  \\
&= \dfrac{1}{2}  \sum_{i, j} \int (K_i -K_i') \, (f_i' f_j'-f_i f_j) \,v_{ij}\,2 \pi b\,db\, d^3 \textbf{v}_{i} d^3 \textbf{v}_{j},\nonumber \\
&= \dfrac{1}{2}  \sum_{j, i} \int (K_j -K_j') \, (f_j' f_i'-f_j f_i) \,v_{ji}\,2 \pi b\,db\, d^3 \textbf{v}_{j} d^3 \textbf{v}_{i},\nonumber\\
&= \dfrac{1}{4}  \sum_{i, j} \int (K_i + K_j -K_i' -K_j') \, (f_i' f_j'-f_i f_j)\nonumber \\
& \hspace{3cm} \times \,v_{ij}\,2 \pi b\,db\, d^3 \textbf{v}_{i} d^3 \textbf{v}_{j}.\nonumber
\end{align}
Eqs.\,\eqref{eq:invariant1} and \eqref{eq:invariant2} are used in Sec.\,\ref{sec:Invariants} to introduce the collisional invariants $K(\textbf{v}) = m, m \,\textbf{v}, m \,v^2 / 2$ which verify $J[K] = 0$.\\

\subsection{Linearized operator}
\label{app:LinColl}

In thermodynamic equilibrium, the velocity distribution functions $f_i(\textbf{v})$ are Maxwellian, equal to $M_i(\textbf{v})$, and the Boltzmann collision integrals vanish due to the conservation of energy in each binary collision, Eq.\,\eqref{eq:binary_momentum_energy}, leading to $M_i' \, M_j' = M_i \, M_j$. Close to equilibrium, the distribution functions are slightly perturbed, of the form $M_i(\textbf{v})\,(1 + \varepsilon\,\phi_i(\textbf{v}))$, where $\varepsilon$ is the Knudsen number (see Sec.\,\ref{sec:Knudsen}), and the $\phi_i$ are solution to the linearized Boltzmann equations, Eq.\,\eqref{eq:eq_phi}. In these equations, the Boltzmann collision integrals $J[f_i, f_j](\textbf{v}_{i})$ reduce to linearized collision integrals $I[\phi_i, \phi_j](\textbf{v}_{i})$
\begin{equation}
J\left[M_i\,(1 + \varepsilon\,\phi_i), M_j\,(1 + \varepsilon\,\phi_j)\right](\textbf{v}_{i}) = \varepsilon\,I[\phi_i,\phi_j](\textbf{v}_{i}),
\end{equation}
at first order in $\varepsilon$, with
\begin{equation}
I[\phi_i,\phi_j](\textbf{v}_i) = \int M_i M_j \, (\phi_i'+\phi_j'-\phi_i-\phi_j) \, v_{ij} \, 2 \pi \, b \, d b \, d^3 \textbf{v}_{j}.
\end{equation}
In the derivations of Sec.\,\ref{sec:solution}, the solutions $\phi_i$ involve vectorial and tensorial fields : $\textbf{K}$  and $\rttensor{K}$. It is then useful to extend, componentwise, the definition of the linearized Boltzmann collision integrals to these fields
\begin{align}
I[\textbf{K}_i,\textbf{K}_j](\textbf{v}_i)  = \int M_i M_j &\, (\textbf{K}_i'+\textbf{K}_j'-\textbf{K}_i-\textbf{K}_j) \nonumber\\
&\times v_{ij} \, 2 \pi \, b \, d b \, d^3 \textbf{v}_{j},
\end{align}
representing a vectorial field, and
\begin{align}
I[\rttensor{K}_{~i},\rttensor{K}_{~j}](\textbf{v}_i)  = \int M_i M_j &\, (\rttensor{K}_{~i}'+\rttensor{K}_{~j}'-\rttensor{K}_{~i}-\rttensor{K}_{~j}) \nonumber\\
&\times v_{ij} \, 2 \pi \, b \, d b \, d^3 \textbf{v}_{j},
\end{align}
representing a tensorial field.

The moments of the collision operator $J$ defined in Sec.\,\ref{app:moments} can also be defined for the linearized operator $I$. They exhibit the same properties with respect to the collisional invariants $K(\textbf{v}) = m, m \,\textbf{v}, m \,v^2 / 2$ with $I[K] = 0$. Of particular interest is the expression of the moment of a function $K(\textbf{v})$ summed over all pairs of species
\begin{align}
I[K] &= \sum_{i, j} \int  K(\textbf{v}_{i}) \, I[\phi_i,\phi_j](\textbf{v}_{i}) \, d^3 \textbf{v}_{i} ,  \\
&= \dfrac{1}{4}  \sum_{i, j} \int (K_i + K_j -K_i' -K_j') \, (\phi_i' + \phi_j'-\phi_i - \phi_j)\nonumber \\
& \hspace{3cm} \times \,v_{ij}\,2 \pi b\,db\, d^3 \textbf{v}_{i} d^3 \textbf{v}_{j}.\nonumber
\end{align}
Indeed, the expression of $(-I[K])$ is a symmetric functional of the functions $K(\textbf{v})$ and $\phi(\textbf{v})$. This is generalized by the definition of the \textit{bracket integrals} $\left[G\times F\right]$
\begin{subequations}
\begin{align}
\left[G \times F\right] &= - \sum_{i,j} \int G_i(\textbf{v}_{i}) \, I[F_i, F_j](\textbf{v}_i) \, d^3 \textbf{v}_i\\
&= \dfrac{1}{4} \sum_{i,j} \int M_i \,M_j\, (G_i'+ G_j' - G_i -G_j) \nonumber\\ 
&\hspace{25mm}\times (F_i' +F_j'-F_i-F_j) \nonumber\\ 
&\hspace{3cm}\times \mathbf{v}_{ij}\, 2 \pi \, b \, d b \, d^3 \textbf{v}_{j} \, d^3 \textbf{v}_i.\nonumber
\end{align}
This functional is symmetric
\begin{equation}
\left[G \times F\right] = \left[F \times G\right].
\end{equation}
It is also a bilinear form
\begin{equation}
\left[G \times (F_1+F_2)\right] = \left[G \times F_1\right]+\left[G \times F_2\right].
\end{equation}
The following generalizations to vectorial and tensorial fields read
\begin{align}
\left[\textbf{G} \cdot \textbf{F}\right] &= - \sum_{i,j} \int \textbf{G}_i(\textbf{v}_{i}) \cdot I[\textbf{F}_i, \textbf{F}_j](\textbf{v}_i) \, d^3 \textbf{v}_i\\
&= \dfrac{1}{4} \sum_{i,j} \int M_i \,M_j\, (\textbf{G}_i'+ \textbf{G}_j' - \textbf{G}_i -\textbf{G}_j) \nonumber\\ 
&\hspace{25mm}\cdot (\textbf{F}_i' +\textbf{F}_j'-\textbf{F}_i-\textbf{F}_j) \nonumber\\ 
&\hspace{3cm}\times \mathbf{v}_{ij}\, 2 \pi \, b \, d b \, d^3 \textbf{v}_{j} \, d^3 \textbf{v}_i.\nonumber
\end{align}
and
\begin{align}
\left[\rttensor{G} : \rttensor{F}\right] &= - \sum_{i,j} \int \rttensor{G}_{~i}(\textbf{v}_{i}) : I\left[\rttensor{F}_{~i}, \rttensor{F}_{~j}\right](\textbf{v}_i) \, d^3 \textbf{v}_i\\
&= \dfrac{1}{4} \sum_{i,j} \int M_i \,M_j\, (\rttensor{G}_{~i}'+ \rttensor{G}_{~j}' - \rttensor{G}_{~i} -\rttensor{G}_{~j}) \nonumber\\ 
&\hspace{25mm}: (\rttensor{F}_{~i}' +\rttensor{F}_{~j}'-\rttensor{F}_{~i}-\rttensor{F}_{~j}) \nonumber\\ 
&\hspace{3cm}\times \mathbf{v}_{ij}\, 2 \pi \, b \, d b \, d^3 \textbf{v}_{j} \, d^3 \textbf{v}_i.\nonumber
\end{align}
\end{subequations}

\subsection{Rotational invariance}
\label{app:Rot}
The general form of the solution to the linearized Boltzmann equation is given in Eq.\,\eqref{eq:phi_grad} using an argument of rotational invariance, that we describe in details in this section. 

Assume that we rotate the reference frame of velocities. We note the functions of vectors with coordinates in the new frame with a tilde, $\tilde f(\textbf{v})$, whereas the same functions in the old frame are noted without tilde, $f(\textbf{v})$. 

If the functions return a scalar, both are related by the matrix of rotation $[R]$, according to
\begin{equation}
\tilde f(\textbf{v}) = f\left([R^{-1}] ~ \textbf{v}\right),
\end{equation}
where $[R^{-1}]$ stands for the inverse rotation.

If the functions return a vector $\textbf{K}$, both are related according to
\begin{equation}
\tilde {\textbf{K}}(\textbf{v}) = [R]~\textbf{K}\left([R^{-1}] ~ \textbf{v}\right).
\end{equation}

First, we want to check that the Boltzmann collision operator $J$ is invariant under this rotation. In the new frame, it reads
\begin{align*}
\tilde J[\tilde f_i, \tilde f_j](\textbf{v}) &= J[f_i, f_j]\left([R^{-1}] ~ \textbf{v}\right)\\
&= \int \left(f_i' f'_j-f_i f_j\right) \, \left|\textbf{v}_{j} - [R^{-1}] ~\textbf{v}\right|\, 2 \pi \, b \, d b \, d^3 \textbf{v}_{j},
\end{align*}
where $f_i = f_i\left([R^{-1}] ~ \textbf{v}\right)$, $f_i'$ and $f_j'$ corresponding to the final collision velocities.
Now, with the change of variable $\textbf{u}_{j} = [R]~ \textbf{v}_{j}$ one gets
\begin{equation}
\tilde J[\tilde f_i, \tilde f_j](\textbf{v}) = J[\tilde f_i, \tilde f_j]\left(\textbf{v}\right),
\end{equation}
since the binary collision is rotationally invariant (see app.\,\ref{app:Collision}).

By the same line of arguments, the linearized Boltzmann collision operator $I$ is also rotationally invariant. Since $I$ is a linear operator, it can be extended to act componentwise on vector fields as well. In these cases, the rotational invariance reads
\begin{equation}
\tilde {{I}}[\tilde {\textbf{K}}_i, \tilde {\textbf{K}}_j](\textbf{v}) = {I}[\tilde {\textbf{K}}_i, \tilde {\textbf{K}}_j]\left(\textbf{v}\right).
\end{equation}
Furthermore, this extension of $I$ to vector fields produces a vector $\textbf{F}$, and the transformation from old to new frame is 
\begin{equation}
\textbf{F} = \tilde I[\tilde {\textbf{K}}_i, \tilde {\textbf{K}}_j](\textbf{v}) = [R]~I[ {\textbf{K}}_i, {\textbf{K}}_j]\left([R^{-1}] ~ \textbf{v}\right).
\end{equation}
Therefore, the rotational invariance implies in this case that
\begin{equation}
\left[R\right]~\textbf{K}_i\left([R^{-1}] ~ \textbf{v}\right) = \textbf{K}_i\left(\textbf{v}\right),
\end{equation}
and the only admissible functional form of $\textbf{K}_i$ is
\begin{equation}
\textbf{K}_i\left(\textbf{v}\right) = K_i(v)~\textbf{v},
\end{equation}
where $K_i(v)$ is a function of the modulus of $\textbf{v}$. 

The same reasoning applies to the componentwise extension of the linearized Boltzmann collision operator $I$ to tensor fields $\rttensor{K}$ leading to the admissible solutions in the form
\begin{equation}
\rttensor{K}_{~i}(\textbf v) = K_i(v) ~ \textbf{v} \otimes \textbf{v},
\end{equation}
when the tensor resulting from the action of $I$ is symmetric, of the form $\textbf{F} \otimes \textbf{F}$.

\section{Momentum and energy equations}
\label{app:fluidEq}

The derivation of the momentum conservation equation proceeds by multiplying the Boltzmann by $m_i \textbf{v}_i$, integrating in velocity, and summing over all the species. Let us focus on the term
\begin{subequations}
\label{eq:momentum}
\begin{align}
&\left[\sum_i  \int d^3\textbf{v} \, m_i \textbf{v} \, \nabla\cdot( \textbf{v} f_i)\right]_\alpha 
= \sum_i  m_i \int d^3\textbf{v} \, v_\alpha \, \nabla_\gamma ( v_\gamma f_i)\nonumber\\
&= \nabla_\gamma \left(\sum_i  m_i \int d^3\textbf{v} \, v_\gamma \, v_\alpha \, f_i \right).
\end{align}
where Einstein's rule of summation has been used (see App.\,\ref{app:Einstein}). Now, remark that $\textbf{v} = \textbf{u} + (\textbf{v} - \textbf{u})$ leads to
\begin{align}
\int d^3\textbf{v}& \, v_\gamma  v_\alpha \, f_i  = \int d^3\textbf{v} \, u_\gamma  u_\alpha \, f_i \\
& + \int d^3\textbf{v} \,  (v_\gamma-u_\gamma)  (v_\alpha-u_\alpha) \, f_i \nonumber\\
& + \int d^3\textbf{v} \,  u_\gamma  (v_\alpha-u_\alpha) \, f_i  + \int d^3\textbf{v} \,  (v_\gamma-u_\gamma) u_\alpha \, f_i \nonumber\\
= ~& n_i \, \left(\textbf{u} \otimes \textbf{u}\right)_{\gamma \alpha}  + \int d^3\textbf{v} \,  (v_\gamma-u_\gamma) \, (v_\alpha-u_\alpha) \, f_i \nonumber\\
&+  u_\gamma \, (n_i \, u_i)_\alpha-n_i \,  u_\gamma \, u_\alpha + ( n_i \, u_i)_\gamma  \,u_\alpha -n_i \, u_\gamma \,u_\alpha.\nonumber
\end{align}
Recall that the fluid velocity $\textbf{u}$ is defined by 
\begin{equation}
\label{eq:FluidV}
\rho\, u_\alpha = \sum_i \rho_i (u_i)_\alpha,
\end{equation}
so that, when summed over the species $i$, the contributions of the last four terms vanish two by two.
Finally introduce the definition of the pressure tensor $\rttensor{\Pi}$ 
\begin{equation}
\label{eq:stress}
\rttensor{\Pi} = \sum_i m_i \int (\textbf{v}-\textbf{u}) \otimes (\textbf{v}-\textbf{u}) \, f_i(t,\textbf{r},\textbf{v}) \, d^3 \textbf{v},
 \end{equation}
\end{subequations}
leads to the momentum conservation equation Eq.\,\eqref{momentum_eq}. 

The energy conservation equation is established by multiplying the Boltzmann equation by $m_i v_i^2 / 2$, integrating in velocity, and summing over all the species. Let us focus on the term 
\begin{subequations}
\label{eq:energy}
\begin{align}
\sum_i \int d^3\textbf{v} \, \dfrac{1}{2} m_i v^2 \, \nabla\cdot( \textbf{v} f_i) & = \sum_i  \dfrac{1}{2} m_i \int d^3\textbf{v} \, v^2 \, \nabla_\gamma ( v_\gamma f_i), \\
& = \nabla_\gamma \left(\sum_i  \dfrac{1}{2} m_i \int d^3\textbf{v} \, v^2 \, v_\gamma \, f_i \right),\nonumber
\end{align}
and remark that $\textbf{v} = \textbf{u} + (\textbf{v} - \textbf{u})$ and $ v^2 = v_\alpha v_\alpha = u_\alpha u_\alpha + 2 u_\alpha (v_\alpha - u_\alpha) + (v_\alpha - u_\alpha)(v_\alpha - u_\alpha)$ leads to
\begin{align}
\int d^3\textbf{v} \, v^2 \, v_\gamma \, f_i  = &\int d^3\textbf{v} \, v^2 \, u_\gamma  f_i \\
&+ \int d^3\textbf{v} \,  u^2 \, (v_\gamma-u_\gamma)  \, f_i\nonumber \\
&+ \int  d^3\textbf{v} \, 2  u_\alpha  (v_\alpha-u_\alpha)\, (v_\gamma-u_\gamma) \, f_i\nonumber \\
&+ \int  d^3\textbf{v} \,  (\textbf{v} - \textbf{u})^2 (v_\gamma-u_\gamma) \, f_i\nonumber .\nonumber
\end{align}
The first term involves the energy density $E$ 
\begin{equation}
E  = \sum_i \int \dfrac{1}{2} m_i v^2 \, f_i(t,\textbf{r},\textbf{v}) \, d^3\textbf{v} = \dfrac{1}{2} \rho u^2 + \dfrac{3}{2} n\, k T.
\end{equation}
The second term leads to a vanishing contribution due to the definition of the fluid velocity, Eq.\,\eqref{eq:FluidV}. The third term involves the stress tensor $\rttensor{\Pi}$, Eq.\,\eqref{eq:stress}. The last term involves the heat flux $\textbf{q}$
\begin{equation}
\textbf{q}  = \sum_i \int \dfrac{1}{2} m_i (\textbf{v}-\textbf{u})^2 (\textbf{v}-\textbf{u}) \, f_i(t,\textbf{r},\textbf{v}) \, d^3 \textbf{v}.
\end{equation}
\end{subequations}
Grouping all the contributions leads to the energy conservation equation \eqref{energy_eq}.

\section{Driving forces}
\label{app:Snider}

The source terms of the linearized Boltzmann equation, Eq.\,\eqref{eq:eq_phi}, are called driving forces since they eventually lead to the expressions of the diffusive fluxes of mass, momentum, and energy, with the appearance of the transport coefficients. We start the derivation of the source terms with the following reformulation
\begin{subequations}
\begin{equation*}
\partial_t M_i + \textbf{v} \cdot \nabla M_i  = M_i \left[ \partial_t \log(M_i) + \textbf{v} \cdot \nabla  \log(M_i)  \right],
\end{equation*}
with
\begin{equation}
\log(M_i) = \log(n_i) - \dfrac{3}{2} \log(T) - \dfrac{m_i (\textbf{v}-\textbf{u})^2}{2 k T} + cste.\\
\end{equation}
Consequently, it follows that
\begin{equation*}
\partial_t \log(M_i) = \partial_t \log(n_i) + \left( \dfrac{m_i c^2}{2 k T}-\dfrac{3}{2}\right)\partial_t \log(T) + \dfrac{m_i}{k T}  \textbf{c} \cdot \partial_t \textbf{u},
\end{equation*}
\begin{align}
\textbf{v}\cdot\nabla \log(M_i) = &\textbf{v}\cdot\nabla \log(n_i)  + \left( \dfrac{m_i c^2}{2 k T}-\dfrac{3}{2}\right)\textbf{v}\cdot\nabla \log(T) \nonumber\\
 & +\dfrac{m_i}{k T}\,\textbf{c} \otimes \textbf{v}  : \nabla \otimes \textbf{u},
\end{align}
\end{subequations}
where we have set $\textbf{c} = \textbf{v} -\textbf{u}$. Let us develop the calculation leading to the last term (see App.\,\ref{app:Einstein} for the notation)
\begin{equation*}
\textbf{v}\cdot\nabla c^2 = v_\alpha \nabla_\alpha (c_\beta c_\beta) = 2\, v_\alpha c_\beta \nabla_\alpha c_\beta = - 2 \, c_\beta v_\alpha \nabla_\alpha u_\beta.
\end{equation*}
Then consider the Euler equations
\begin{subequations}
\begin{align}
&\partial_t \rho_i + \nabla . \left( \rho_i \textbf{u} \right) = 0, \\
&\partial_t (\rho \textbf{u}) + \nabla.\left( \rho \textbf{u} \otimes \textbf{u} + p \,\rttensor{Id} \right)=0,  \\
&\partial_t E + \nabla. \left( (E + p) \textbf{u} \right)=0,  
\end{align}
under this useful form
\begin{align}
& \partial_t n_i + \nabla( n_i \textbf{u} ) = 0, \\
&\partial_t \textbf{u} + (\textbf{u} . \nabla) \textbf{u} + \dfrac{\nabla p}{\rho} = 0, \\
&\partial_t (k T) + \textbf{u} . \nabla (k T) + \dfrac{2 p}{3n}\nabla . \textbf{u}=0. 
\end{align}
The last equation for the time evolution of temperature is easily derived once the following equation for the time evolution of kinetic energy is established
\begin{equation}
\partial_t \left(\dfrac{1}{2} \rho u^2\right) + \nabla \cdot \left(\dfrac{1}{2} \rho u^2\, \textbf{u}\right) = - \textbf{u} \cdot \nabla p.
\end{equation}
\end{subequations}
Substituting the time derivatives directly leads to the driving forces term expression
\begin{subequations}
\begin{align}
\dfrac{1}{M_i} ( \partial_t M_i & + \textbf{v} \cdot \nabla M_i ) \nonumber\\
& = \textbf{c}.\dfrac{\nabla n_i}{n_i}  - \nabla.\textbf{u} \, +  \left( \dfrac{m_i c^2}{2 k T}- \dfrac{3}{2}\right) \textbf{c}.\dfrac{\nabla T}{T} \nonumber\\
& - \left(\dfrac{m_i c^2}{2 k T}-\dfrac{3}{2}\right) \dfrac{2 p}{3 n k T} \nabla . \textbf{u}\nonumber \\
& + \dfrac{m_i}{k T}\, \textbf{c} \otimes \textbf{c} : \nabla \otimes \textbf{u}\, -\dfrac{m_i}{k T} \dfrac{\textbf{c}.\nabla p}{\rho},
\end{align}
which gives in the case of a ideal gas ($p=n k T$)
\begin{align}
\dfrac{1}{M_i} \left(\partial_t M_i + \textbf{v} \cdot \nabla M_i\right) & =  \textbf{c} \cdot \left( \dfrac{\nabla n_i}{n_i}-\dfrac{m_i}{m} \dfrac{\nabla n}{n} \right) \\
& + \left( \dfrac{m_i c^2}{2T}- \dfrac{3}{2}-\dfrac{m_i}{m}\right) \textbf{c}\cdot\dfrac{\nabla T}{T}  \nonumber\\
& - \dfrac{m_i c^2}{3 k T}\, \nabla\cdot\textbf{u} + \dfrac{m_i}{k T}\, \textbf{c} \otimes \textbf{c} : \nabla \otimes \textbf{u}.\nonumber
\end{align}
Finally, recall that $n = \sum_k n_k$, and evidence that the symmetric traceless tensor $(\textbf{c} \otimes \textbf{c} - \dfrac{1}{3} c^2 \rttensor{Id})$ acts on $(\nabla \otimes \textbf{u})$ so that only the symmetric traceless part $(\rttensor S)$ of the tensor $(\nabla \otimes \textbf{u})$ contributes to the tensorial contraction
\begin{equation}
(\rttensor{S})_{\alpha \beta} = \dfrac{1}{2} \,\left(\nabla_\alpha u_\beta + \nabla_\beta u_\alpha\right) - \dfrac{1}{3} \, (\nabla \cdot \textbf{u}) ~ \delta_{\alpha \beta}.
\end{equation}
Likewise, when the tensor $\rttensor{S}$ is used, the contraction with $\textbf{c} \otimes \textbf{c}$ gives the same result. This little trick is useful to get more direct derivation
\begin{align}
\dfrac{1}{M_i}\,(\partial_t M_i + \textbf{v} \cdot \nabla M_i)
&= \left( \dfrac{m_i \textbf{c}^2}{2 k T}- \dfrac{3}{2}-\dfrac{m_i}{m}\right) \, \textbf{c} \cdot \nabla \log T \nonumber\\
&+ \dfrac{m_i}{k T} ~ \textbf{c} \otimes \textbf{c} : \rttensor{S} \\
&+\dfrac{1}{n_i}\,\sum_k \left( \delta_{ik} - \dfrac{m_i n_i}{\rho}\right)\, \textbf{c} \cdot \nabla n_k,\nonumber
\end{align}
\end{subequations}

\section{Bracket integrals}
\label{app:TransportCoeff}

The rate of entropy production discussed in Sec.\,\ref{sec:Entropy} involves bracket integrals, that are developed in the following.

\subsection{Temperature gradient}

Related to the presence of a temperature gradient only, the rate of entropy production involves the solution $\phi^T$ in the bracket integral
\begin{align}
&\left[\phi^T \times \phi^T \right] = - \sum_{i,j} \int \phi^T_i(\textbf{c}_{i}) \, I[\phi^T _i, \phi^T _j](\textbf{c}_i) \, d^3 \textbf{c}_i\\
&= -\sum_{i} \int \phi^T_i(\textbf{c}_{i}) \, M_i\, \left( \dfrac{m_i c_i^2}{2 k T}- \dfrac{3}{2}-\dfrac{m_i}{m}\right) \, \textbf{c}_i \cdot \nabla \log T \, d^3 \textbf{c}_i\nonumber
\end{align}
using Eqs.\,\eqref{eq:phiT}. Now, $\phi^T(\textbf{c}) = - K^T(c)~\textbf{c}\cdot\nabla \log T$ is a scalar. Therefore
\begin{equation}
\left[\phi^T \times \phi^T \right]= \sum_{i} \textbf{V}_i \cdot \nabla \log T,
\end{equation}
with 
\begin{align}
\textbf{V}_i&= - \int \phi^T_i(\textbf{c}_{i}) \, M_i\, \left( \dfrac{m_i c_i^2}{2 k T}- \dfrac{3}{2}-\dfrac{m_i}{m}\right) \, \textbf{c}_i \, d^3 \textbf{c}_i\nonumber\\
&= \int \textbf{c}_i ~ K^T_i(c_i)\, M_i\, \left( \dfrac{m_i c_i^2}{2 k T}- \dfrac{3}{2}-\dfrac{m_i}{m}\right)  ~\textbf{c}_i\cdot\nabla \log T  \, d^3 \textbf{c}_i\nonumber\\
&=\dfrac{1}{3} ~ V_i~ \nabla \log T
\end{align}
using Eq.\,\eqref{eq:IntegralVector}, with
\begin{align}
V_i&= \int K^T_i(c_i)~\textbf{c}_i\cdot\textbf{c}_i \, M_i\, \left( \dfrac{m_i c_i^2}{2 k T}- \dfrac{3}{2}-\dfrac{m_i}{m}\right)  \, d^3 \textbf{c}_i,\nonumber\\
&= \sum_j \int K^T_i(c_i)~\textbf{c}_i\cdot I[\textbf{K}^T _i, \textbf{K}^T _j](\textbf{c}_i)\, d^3 \textbf{c}_i,\nonumber\\
&= -\sum_j \int \textbf{K}^T_i(\textbf{c}_i)\cdot I[\textbf{K}^T _i, \textbf{K}^T _j](\textbf{c}_i)\, d^3 \textbf{c}_i,
\end{align}
using eq.\,\eqref{eq:KTi}. Finally, one gets
\begin{align}
\left[\phi^T \times \phi^T \right] &= \dfrac{1}{3} \sum_{i} V_i |\nabla \log T|^2 \nonumber\\
&= \dfrac{1}{3} \left[\textbf{K}^T \cdot \textbf{K}^T \right] |\nabla \log T|^2.
\end{align}

\subsection{Velocity gradients}

Related to the presence of velocity gradients only, the rate of entropy production involves the solution $\phi^S$ in the bracket integral
\begin{align}
 \left[\phi^S \times \phi^S \right] &=- \sum_{i,j} \int \phi^S_i(\textbf{c}_{i}) \, I[\phi^S _i, \phi^S _j](\textbf{c}_i) \, d^3 \textbf{c}_i\\
&= -\sum_{i} \int \phi^S_i(\textbf{c}_{i}) \, M_i\, \dfrac{m_i}{k T} \, \textbf{c}_i \otimes \textbf{c}_i : \rttensor{S}\, d^3 \textbf{c}_i\nonumber\\
&=\sum_i \rttensor{T}_{~i} : \rttensor S,\nonumber
\end{align}
with
\begin{align}
\rttensor{T}_{~i}&= - \int \phi^S_i(\textbf{c}_{i}) \, M_i\, \dfrac{m_i}{k T} \, \textbf{c}_i \otimes \textbf{c}_i \, d^3 \textbf{c}_i\nonumber\\
&= \int  \textbf{c}_i \otimes \textbf{c}_i \,K^S_i(c_i) \, M_i\, \dfrac{m_i}{k T} \, \left(\textbf{c}_i \otimes \textbf{c}_i  : \rttensor{S}\right)  \, d^3 \textbf{c}_i\nonumber\\
&=\dfrac{2}{15} ~ T_i~ \rttensor{S},
\end{align}
using Eq.\,\eqref{eq:IntegralTensor},
with
\begin{align}
T_i&= \int K^S_i(c_i)~\textbf{c}_i \otimes \textbf{c}_i  :\textbf{c}_i \otimes \textbf{c}_i  \, M_i\,  \dfrac{m_i}{k T} \, d^3 \textbf{c}_i,\nonumber\\
&= \sum_j \int K^S_i(c_i)~\textbf{c}_i \otimes \textbf{c}_i  : I\left[\rttensor{K}^S_{~i}, \rttensor{K}^S_{~j}\right](\textbf{c}_i)\, d^3 \textbf{c}_i\nonumber\\
&= -\sum_j \int \rttensor{K}^S_{~i}(\textbf{c}_i) : I\left[\rttensor{K}^S_{~i}, \rttensor{K}^S_{~j}\right](\textbf{c}_i)\, d^3 \textbf{c}_i,
\end{align}
using Eq.\,\eqref{eq:KSi}. Finally, one gets
\begin{align}
\left[\phi^S \times \phi^S \right] &= \dfrac{2}{15} \sum_{i} T_i ~ \rttensor{S} : \rttensor{S}\nonumber\\
&= \dfrac{2}{15} \, \left[\rttensor{K}^S : \rttensor{K}^S\right]\, \rttensor S : \rttensor S.
\end{align}

\subsection{Partial density gradients}

Related to the presence of gradients of partial densities only, the rate of entropy production involves the solutions $\phi^k$ and $\phi^l$ in the bracket integral
\begin{align}
&\left[\phi^k \times \phi^l \right] = - \sum_{i,j} \int \phi^k_i(\textbf{c}_{i}) \, I[\phi^l _i, \phi^l _j](\textbf{c}_i) \, d^3 \textbf{c}_i\\
&= -\sum_{i} \int \phi^k_i(\textbf{c}_{i}) \, M_i\, \dfrac{1}{n_i}\, \left( \delta_{il} - \dfrac{m_i n_i}{\rho}\right)\, \textbf{c}_i \cdot \nabla n_l \, d^3 \textbf{c}_i\nonumber\\
&= \sum_{i} \textbf{V}_i \cdot \nabla n_l,\nonumber
\end{align}
with
\begin{align}
\textbf{V}_i&= - \int \phi^k_i(\textbf{c}_{i}) \, M_i\, \dfrac{1}{n_i}\, \left( \delta_{il} - \dfrac{m_i n_i}{\rho}\right)\, \textbf{c}_i \, d^3 \textbf{c}_i\nonumber\\
&= \int \textbf{c}_i ~ K^k_i(c_i)\,  M_i\, \dfrac{1}{n_i}\, \left( \delta_{il} - \dfrac{m_i n_i}{\rho}\right)  ~\textbf{c}_i\cdot\nabla n_k  \, d^3 \textbf{c}_i\nonumber\\
&=\dfrac{1}{3} ~ V_i~ \nabla n_k
\end{align}
using Eq.\,\eqref{eq:IntegralVector}, with
\begin{align}
V_i&= \int K^k_i(c_i)~\textbf{c}_i\cdot\textbf{c}_i \, M_i\, \dfrac{1}{n_i}\, \left( \delta_{il} - \dfrac{m_i n_i}{\rho}\right)   \, d^3 \textbf{c}_i,\nonumber\\
&= \sum_j \int K^k_i(c_i)~\textbf{c}_i\cdot I[\textbf{K}^l_i, \textbf{K}^l_j](\textbf{c}_i)\, d^3 \textbf{c}_i,\nonumber\\
&= -\sum_j \int \textbf{K}^k_i(\textbf{c}_i)\cdot I[\textbf{K}^l _i, \textbf{K}^l _j](\textbf{c}_i)\, d^3 \textbf{c}_i,
\end{align}
using eq.\,\eqref{eq:Kij}. Finally, one gets
\begin{align}
\left[\phi^k \times \phi^l \right] &= \dfrac{1}{3} \sum_{i} V_i ~ \nabla n_k  \cdot \nabla n_l \nonumber\\
&= \dfrac{1}{3} \left[\textbf{K}^k \cdot \textbf{K}^l \right]  \nabla n_k  \cdot \nabla n_l.
\end{align}

\subsection{Temperature and density gradients}

Related to the presence of simultaneous gradients of temperature and partial density only, the rate of entropy production involves the solutions $\phi^T$ and $\phi^k$ in the bracket integral
\begin{align}
&\left[\phi^T \times \phi^k \right] = - \sum_{i,j} \int \phi^T_i(\textbf{c}_{i}) \, I[\phi^k _i, \phi^k _j](\textbf{c}_i) \, d^3 \textbf{c}_i\\
&= -\sum_{i} \int \phi^T_i(\textbf{c}_{i}) \, M_i\, \dfrac{1}{n_i}\, \left( \delta_{ik} - \dfrac{m_i n_i}{\rho}\right)\, \textbf{c}_i \cdot \nabla n_k \, d^3 \textbf{c}_i\nonumber\\
&= \sum_{i} \textbf{V}_i \cdot \nabla n_k,\nonumber
\end{align}
with
\begin{align}
\textbf{V}_i&= - \int \phi^T_i(\textbf{c}_{i}) \, M_i\, \dfrac{1}{n_i}\, \left( \delta_{ik} - \dfrac{m_i n_i}{\rho}\right)\, \textbf{c}_i \, d^3 \textbf{c}_i\nonumber\\
&= \int \textbf{c}_i ~ K^T_i(c_i)\,  M_i\, \dfrac{1}{n_i}\, \left( \delta_{ik} - \dfrac{m_i n_i}{\rho}\right)  ~\textbf{c}_i\cdot\nabla \log T  \, d^3 \textbf{c}_i\nonumber\\
&=\dfrac{1}{3} ~ V_i~ \nabla \log T
\end{align}
using Eq.\,\eqref{eq:IntegralVector}, with
\begin{align}
V_i&= \int K^T_i(c_i)~\textbf{c}_i\cdot\textbf{c}_i \, M_i\, \dfrac{1}{n_i}\, \left( \delta_{ik} - \dfrac{m_i n_i}{\rho}\right)   \, d^3 \textbf{c}_i,\nonumber\\
&= \sum_j \int K^T_i(c_i)~\textbf{c}_i\cdot I[\textbf{K}^k_i, \textbf{K}^k_j](\textbf{c}_i)\, d^3 \textbf{c}_i,\nonumber\\
&= -\sum_j \int \textbf{K}^T_i(\textbf{c}_i)\cdot I[\textbf{K}^k _i, \textbf{K}^k _j](\textbf{c}_i)\, d^3 \textbf{c}_i,
\end{align}
using eq.\,\eqref{eq:Kij}. Finally, one gets
\begin{align}
\left[\phi^T \times \phi^k \right] &= \dfrac{1}{3} \sum_{i} V_i ~ \nabla \log T  \cdot \nabla n_k \nonumber\\
&= \dfrac{1}{3} \left[\textbf{K}^T \cdot \textbf{K}^k \right]  \nabla \log T  \cdot \nabla n_k.
\end{align}

\section{Tensor integrals}

In order to clarify the derivation of the transport coefficients expressions, several integral terms calculations are now presented. Consider an arbitrary function $F(c)$ of the modulus of $\textbf{c}$, a constant vector $\textbf{a}$, and a constant traceless symmetric tensor $\rttensor S$, we need to evaluate integrals over the following vector forms
\begin{align}
\label{eq:IntegralVector}
\textbf{V} & = \int d^3\textbf{c} \,\,\textbf{c} \,\, F(c) \,\, \textbf{c} \cdot \textbf{a},\\
 &= \dfrac{1}{3} \, \textbf{a} \int F(c) \, c^2 \,  d^3\textbf{c} \nonumber\\
&= \dfrac{4 \pi}{3} \, \textbf{a} \int_0^{\tiny\infty} F(c) \, c^4 \,  dc, \nonumber
\end{align}
and
\begin{align}
\label{eq:IntegralTensor}
\rttensor{T} & = \int d^3\textbf{c} \,\, \textbf{c} \otimes \textbf{c} \,\, F(c) \,\, \textbf{c} \otimes \textbf{c} : \rttensor S\\
&= \dfrac{2}{15} ~ \rttensor{S} \int F(c) \, c^4 \,  d^3\textbf{c} \nonumber\\
&= \dfrac{8 \pi}{15} ~ \rttensor{S} \int_0^{\tiny\infty} F(c) \, c^6 \,  dc.\nonumber
\end{align}

We first address Eq.\,\eqref{eq:IntegralVector}. Let us assume the vector $\textbf{a}$ along the $x$ direction
\begin{align}
&V_x = a ~\int_{-\infty}^{+\infty} dc_x \, dc_y \, dc_z ~ c_x^2 ~ F(c), \\
&V_y = a ~\int_{-\infty}^{+\infty} dc_x \, dc_y \, dc_z ~  c_x c_y ~ F(c), \nonumber \\
&V_z = a ~\int_{-\infty}^{+\infty} dc_x \, dc_y \, dc_z ~ c_x c_z ~ F(c). \nonumber
\end{align}
Then, $V_y = V_z = 0$ as integrals of an odd function of $c_x$, in particular. Finally, by a change of variables from $c_x$ to $c_y$ or $c_z$, one gets
\begin{align}
\int_{-\infty}^{+\infty} dc_x \, dc_y \, dc_z ~ c_x^2 ~ F(c) &= \int_{-\infty}^{+\infty} dc_x \, dc_y \, dc_z ~ c_y^2 ~ F(c) \nonumber\\
&= \int_{-\infty}^{+\infty} dc_x \, dc_y \, dc_z ~ c_z^2 ~ F(c) \nonumber \\
&= \dfrac{1}{3} \int  c^2 ~ F(c) \, d^3\textbf{c}.
\end{align}

The case of Eq.\,\eqref{eq:IntegralTensor} is more burdensome and needs to introduce spherical coordinates : $c_x = c \sin\theta \cos\phi, c_y = c \sin\theta \sin\phi, c_z = c \cos\theta$ with $d^3 c = c^2 \sin\theta \,dc\, d\theta\, d\phi$.
Using Einstein notations of App.\,\ref{app:Einstein} the components of $\rttensor{T}$ write
\begin{equation*}
T_{i j} = \int d^3\textbf{c} \,\, c_i \, c_j \,\, F(c) \,\, c_\alpha c_\beta \, S_{\beta \alpha}.
\end{equation*}

\begin{itemize}
\item Case $i \neq j$: ($T_{xy}$; $T_{xz}$; $T_{yz}$)
\end{itemize}
All the terms other than $(\alpha = i,\, \beta = j)$ and $(\alpha = j,\, \beta = i)$ vanishes as integrals over odd functions of the velocity components. Since $\rttensor{S}$ is symmetric
\begin{equation*}
T_{i j} =  2\, S_{ij} \int d^3\textbf{c} \,\, c_i^2 \, c_j^2 \,\, F(c)
\end{equation*}
and since $c_x$, $c_y$, and $c_z$, are integration variables and $F(c)$ is invariant over any interchange of $c_x$, $c_y$, and $c_z$, the three components are equal to
\begin{align*}
T_{ij} & = 2\, S_{ij} \int d^3\textbf{c} \,\, c_x^2 \, c_y^2 \,\, F(c), \\
&= 2\, S_{ij} \int_0^{\tiny \infty} c^2\, dc \,\, c^4 \,\, F(c) \int_0^{2\pi}  d\phi \,\cos^2\phi \sin^2\phi \\
& \hspace{4cm} \times \int_0^\pi \sin(\theta) \, d\theta \,\sin^4\theta, \\
&= S_{ij} \,\dfrac{8 \pi}{15} \int_0^{\tiny \infty} dc \,\, c^6 \,\, F(c) = S_{ij}\,\dfrac{2}{15} \int d^3\textbf{c} \,\, c^4 \,\, F(c).
\end{align*}

\begin{itemize}
\item Case $i = j$: ($T_{xx}$; $T_{yy}$; $T_{zz}$)
\end{itemize}
All the terms other than $(\alpha = \beta)$ vanishes as integrals over odd functions of the velocity components
\begin{align*}
T_{i i} &= S_{ii} \int d^3\textbf{c} \,\, c_i^4 \,\, F(c) + \sum_{\alpha \neq i} S_{\alpha \alpha} \int d^3\textbf{c} \,\, c_i^2 \, c_\alpha^2 \,\, F(c),\\
&= S_{ii} \int d^3\textbf{c} \,\, c_i^4 \,\, F(c) + \sum_{\alpha \neq i} S_{\alpha \alpha} \dfrac{1}{15} \int d^3\textbf{c} \,\, c^4 \,\, F(c),\\
&= S_{ii} \int d^3\textbf{c} \,\, c_i^4 \,\, F(c) - S_{ii} \dfrac{1}{15} \int d^3\textbf{c} \,\, c^4 \,\, F(c),
\end{align*}
with the second integral already worked out in the case $i \neq j$
and using the fact that $\rttensor S$ is traceless. Since $c_x$, $c_y$, and $c_z$, are integration variables and $F(c)$ is invariant over any interchange of $c_x$, $c_y$, and $c_z$,  the first integral can be evaluated using spherical coordinates with $c_i = c_x$
\begin{align*}
\int d^3\textbf{c} \,\, c_i^4 \,\, F(c) &= \int d^3\textbf{c} \,\, c_z^4 \,\, F(c),\\
&= \int_0^{\tiny \infty} c^2\, dc \,\, c^4 \,\, F(c) 
\times \int_0^{2\pi}  d\phi,  \\
&\times \int_0^\pi \sin(\theta) \, d\theta \,\cos^4\theta, \\
&= \dfrac{1}{5}\int d^3\textbf{c} \,\, c^4 \,\, F(c).
\end{align*}
With this last result, Eq.\,\eqref{eq:IntegralTensor} is proven.

\section{Einstein rules of summation}
\label{app:Einstein}

In this section are introduced Einstein's rule of summation over repeated indexes, for instance
\begin{equation*}
\textbf{v} = v_\alpha \textbf{e}_\alpha = v_x \textbf{e}_x + v_y \textbf{e}_y + v_z \textbf{e}_z, 
\end{equation*}
with tensor product 
\begin{equation*}
\left(\textbf{u} \otimes \textbf{v}\right)_{\alpha \beta} = u_\alpha \, v_\beta, 
\end{equation*}
and the scalar products applied to tensors
\begin{equation*}
\left(\textbf{u} \cdot \rttensor{T}\right)_{\alpha} = u_\gamma \, T_{\gamma \alpha}, 
\end{equation*}
\begin{equation*}
\rttensor{T} : \rttensor{W} = T_{\alpha \beta} \, W_{\beta \alpha}. 
\end{equation*}

%
\end{document}